\documentclass[12pt,linesnumbered,ruled,vlined]{article}
\usepackage{mathptmx}
\usepackage[utf8x]{inputenc}
\usepackage{color}
\usepackage[table,xcdraw]{xcolor}
\usepackage{tabularx}
\usepackage{natbib}
\usepackage{array}
\usepackage{booktabs}
\usepackage{mathtools}
\usepackage{amsmath}
\usepackage{amsthm}
\usepackage{amssymb}
\usepackage{graphicx}
\usepackage{multirow}
\usepackage{lscape} 
\usepackage{makecell} 
\usepackage[unicode=true,pdfusetitle,
 bookmarks=true,bookmarksnumbered=false,bookmarksopen=false,
 breaklinks=false,pdfborder={0 0 1},backref=false,colorlinks=true]
 {hyperref}
\hypersetup{
 pdfborderstyle={},pdfborderstyle={},pdfborderstyle={},urlcolor=blue,linkcolor=blue,citecolor=blue}

\makeatletter

\@ifundefined{date}{}{\date{}}

\usepackage{amsfonts}
\usepackage{dsfont}

\numberwithin{equation}{section}

\allowdisplaybreaks[4]
\DeclareMathAlphabet{\mathcal}{OMS}{cmsy}{m}{n}

\makeatother

\usepackage{authblk}
\usepackage{tikz}

\begin{document}


\title{Machine Learning for Economic Forecasting: An Application to China's GDP Growth}
\author[1,2]{Yanqing Yang\thanks{yanqingyang@fudan.edu.cn}}
\author[1]{Xingcheng Xu\thanks{xingcheng.xu18@gmail.com}}
\author[1]{Jinfeng Ge\thanks{gejinfeng@pjlab.org.cn}}
\author[1]{Yan Xu\thanks{xuyan@pjlab.org.cn}}
\affil[1]{Shanghai Artificial Intelligence Laboratory}
\affil[2]{Fudan University}

\date{July 4, 2024}

\maketitle


\begin{abstract}
\vspace{-0.1cm}
This paper aims to explore the application of machine learning in forecasting Chinese macroeconomic variables. Specifically, it employs various machine learning models to predict the quarterly real GDP growth of China, and analyzes the factors contributing to the performance differences among these models. Our findings indicate that the average forecast errors of machine learning models are generally lower than those of traditional econometric models or expert forecasts, particularly in periods of economic stability. However, during certain inflection points, although machine learning models still outperform traditional econometric models, expert forecasts may exhibit greater accuracy in some instances due to experts' more comprehensive understanding of the macroeconomic environment and real-time economic variables. In addition to macroeconomic forecasting, this paper employs interpretable machine learning methods to identify the key attributive variables from different machine learning models, aiming to enhance the understanding and evaluation of their contributions to macroeconomic fluctuations.

\end{abstract}
	\vspace{-0.3cm}
\smallskip
	
 \smallskip


\section{Introduction}
As China's economy enters the ``new normal", changes in growth momentum and global political, economic, and financial environments have increased the difficulty of macroeconomic forecasting based on structural modeling. Exogenous shocks, such as epidemics and conflicts, have also contributed to a decline in forecasting accuracy. Concurrently, advancements in big data and artificial intelligence algorithms have introduced new tools and methods for macroeconomic forecasting and policy regulation. In this paper, we employ cutting-edge machine learning algorithms to forecast China's macroeconomy and examine the interpretability of these forecasting methods.

This paper focuses on forecasting China's GDP. GDP is the core indicator of national economic accounting and the primary basis for assessing macroeconomic performance and formulating economic policies. In recent years, both international and domestic macroeconomic forecasting have advanced significantly. The continuous development of econometric models and improved data availability have facilitated research on forecasting China's GDP growth (\citealp{QuanBing2017}; \citealp{ChenLei2018}; \citealp{ZhangJingFan2018}; \citealp{FeiZhaoQi2019}; \citealp{LiangFang2021}). Meanwhile, the rapid development of artificial intelligence algorithms and the emergence of high-frequency data have positioned machine learning as a crucial emerging tool in economic forecasting (\citealp{belloni2014high}). International research organizations and academics are increasingly applying machine learning models in macroeconomic forecasting. Existing international research indicates that machine learning models generally outperform traditional econometric models in predictive accuracy (\citealp{bajari2015machine}; \citealp{richardson2021nowcasting}).

The application of machine learning models in macroeconomic forecasting in China remains at an early stage. To explore the use of machine learning models for forecasting China's macroeconomy and to assess the effectiveness of different forecasting methods, this paper employs various machine learning models to forecast China's GDP growth rate. These models include machine learning models, combined models of machine learning and econometric method, and multiple models (e.g., mean outcomes of combined models, weighted outcomes of machine leaning models, etc.).

This paper compares the prediction results of machine learning models with traditional econometric models and expert predictions. It further analyzes and explores the predictive performance of various models and the reasons for their effectiveness. The comparison spans different periods, focusing on the effects and causes of economic fluctuations and inflection points. Additionally, this paper employs methods of interpretable machine learning to assess the importance of different variables to the forecasting results from both global and local perspectives. It tries to explain drivers of differences in model forecasting and performance.

The empirical results in this paper indicate that machine learning models, machine learning-econometric combined models, or mixed models generally exhibit higher forecasting accuracy than traditional econometric models. During periods of economic stability and minor fluctuations, machine learning models and their combinations are more accurate than econometric models and expert forecasts. However, at historical inflection points, when economic fluctuations exceed the historical range of the training data, most machine learning models can predict these inflection points, but their accuracy is typically lower than that of expert predictions.

The rest of the paper is organized as follows: Section 2 reviews relevant literature, particularly the application of artificial intelligence methods in macroeconomic forecasting. Section 3 introduces the data and data processing procedures. Section 4 describes the forecasting models and the methods used for combining models and out-of-sample evaluation. Section 5 compares the forecasting results of different models across various periods. Section 6 analyzes the interpretability of the models. Section 7 presents the robustness analysis. Section 8 provides the conclusion.

\section{Literature Review}
\subsection{Forecasts of the Chinese Economy}
In the literature on forecasting macroeconomic variables in China, econometric model forecasting is the predominant method. Traditional econometric models fall into several major approaches. The first one is vector autoregressive (VAR) models, which capture complex dynamic relationships between variables (\citealp{FengSiXian2012}). The second apporach is the mixed-frequency data (MIDAS) models and their derivatives, which enhance forecasting accuracy by using high-frequency data to predict low-frequency economic variables. For instance, \cite{FeiZhaoQi2019} used mixed-frequency techniques to construct a daily prior index for short-term fluctuations in China's macroeconomy, improving predictions of economic fluctuations. The third apporach is mixed-frequency vector autoregression (MF-VAR) models. \cite{ZhangJingFan2018} developed a Bayesian estimation-based MF-BVAR model, finding it significantly more accurate in forecasting core macroeconomic variables such as GDP, CPI, and PPI compared to the homoskedastic VAR and MIDAS models. The fourth approach is autoregressive integrated moving average (ARIMA) models, which uses historical time-series data to forecast future values and is widely applied in economic and financial forecasting (\citealp{ZhanJianZhi2009x11}).

Given that the accuracy of econometric models is inevitably influenced by future events, such as national economic policies, some studies have incorporated expert research in forecasting China's macroeconomic variables. Other studies have combined expert insights with model forecasting to leverage the strengths of both approaches and enhance forecasting accuracy. \cite{LiangFang2021} constructed a macroeconomic system with 15 domestic variables, used multidimensional high-frequency macroeconomic data to forecast GDP growth through mixed-frequency vector autoregressive modeling, and compared these forecasts with expert predictions, which is from the ``Langrun Forecasts".

\subsection{Machine Learning Models}
In recent years, research on predicting macroeconomic variables using machine learning models has proliferated. Several studies have employed machine learning to forecast macroeconomic data and compared their results with traditional methods. For instance, \cite{bajari2015machine} utilized machine learning methods to forecast consumer demand and found these methods to be more accurate than traditional panel data and logistic models for out-of-sample forecasts.

The integration of machine learning into real-time macroeconomic forecasting can enhance forecasting accuracy. In real-time macroeconomic forecasting, Dynamic Factor Models (DFMs) have been widely adopted by institutions worldwide. \cite{giannone2008nowcasting} pioneered the application of DFM models for real-time GDP forecasting, aiding the Federal Reserve's policy formulation. Subsequently, central banks and international organizations, such as the ECB (\citealp{marozzi2021ecb}) and the World Bank (\citealp{dauphin2022nowcasting}), have developed their own DFM models. \cite{richardson2021nowcasting} utilized a machine learning model to process a database of over 600 variables, yielding real-time GDP growth forecasts for New Zealand that outperformed traditional AR and DFM models. The machine learning model demonstrated robust performance during the COVID-19 pandemic. \cite{woloszko2020tracking} used Google data to construct weekly GDP growth rates for OECD countries, with the machine learning models' average error being 17 percentage points lower than that of the traditional AR model for both OECD and G20 countries.

\subsection{Interpretable Machine Learning}
Machine learning algorithms typically employ a nonlinear and nonparametric approach, where model complexity is controlled by hyperparameters selected through cross-validation. While these models leverage nonlinearities in macroeconomic predictions to enhance accuracy, their nonlinear relationships are not easily visualized or understood, creating a ``black-box" challenge. To address this issue, interpretable machine learning methods aim to resolve the opacity of these models.

Interpretable machine learning studies for macro-prediction can be categorized into three main areas: model interpretability based on cognitive psychology (\citealp{miller2019explanation}), technical approaches to machine learning interpretability (\citealp{doshi2017towards}), and discussions grounded in econometric and statistical methods (\citealp{chernozhukov2018double}). \cite{miller2019explanation} argues that traditional econometric or statistical methods are not inherently more explanatory than machine learning. Linear models trained on abstract features, such as those obtained via principal component analysis, similarly lack clear economic interpretation.

With respect to the latter two main areas, interpretability methods primarily focus on assessing the importance of input variables for prediction. Commonly used ``global method" include tree model-based ranked with Gini importance (\citealp{breiman2001random}). Commonly used ``local mehtods" include LIME method (\citealp{ribeiro2016should}) and DeepLIFT method (\citealp{shrikumar2017learning}). In addition, \cite{lundberg2017unified} demonstrate that the Shapley value approach provides a unified framework, integrating both global and local methods. It also ensures consistency between global and local approaches.

\subsection{Economic Volatility and Inflection Points}
Accurately predicting inflection points in GDP growth over the economic cycle using high-frequency data is a key goal for economists and research institutes, though it remains challenging.  \cite{ho2023forecasting} evaluates macroeconomic forecasts during the COVID-19 pandemic, highlighting the importance of including additional real-time information due to the unpredictability of such events and their unique economic shocks, which are historically rare and distinct from ordinary economic cycles. Understanding what information to include in forecasts requires a deep comprehension of economic shocks and macroeconomic mechanisms, making pre-modeling data selection more difficult. Thus, understanding the drivers behind data fluctuations is crucial.

The COVID-19 pandemic provides a valuable example for macroeconomic forecasting research. Data from outside the economic domain, such as public health data, became crucial during the pandemic. The rapid changes during this period increased the demands on forecasts from economists and state agencies. Quarterly or even monthly forecasts proved insufficient to meet policy needs during the pandemic. Forecasting economic fluctuations and macroeconomic trends during special periods, such as pandemics, is an important direction for future research (\citealp{liu2021panel}).

\section{Data}
In this paper, we forecast China's quarterly GDP growth rate. Based on existing studies (\citealp{ZhangJingFan2018}; \citealp{LiangFang2021}), we employ about 20 macroeconomic variables in the model, such as industrial value added, purchasing managers' index (PMI), retail sales of consumer goods, freight transportation, steel production, power generation, and GDP growth rates of other major economies. Variable names, categories, data frequencies and sampling periods are presented in Table \ref{tab:macro-variable}. The data on China were sourced from National Bureau of Statistic of China, and GDP of other economies were sourced from their official statistical bureaus.
    
\begin{table}
    \small
    \centering
     \begin{tabularx}{\textwidth}{l l l l}
        \toprule
        \textbf{Variable Name} & \textbf{Category} & \textbf{Frequency} & \textbf{Sample Period} \\
        \midrule
        China GDP & GDP & Quarterly YoY & 1992Q1-2023Q4 \\
        Floor Space of New Buildings & Real Estate & Monthly YoY & 1999Q1-2023Q4 \\
        Floor Space of Commercial Buildings Sold & Real Estate & Monthly YoY & 1999Q1-2023Q4 \\
        Imports & Import and Export & Monthly YoY & 1995Q1-2023Q4 \\
        Exports & Import and Export & Monthly YoY & 1995Q1-2023Q4 \\
        Industrial Added Value & Production & Monthly YoY & 1995Q1-2023Q4 \\
        PMI: Manufacturing & Production & Monthly & 2005Q1-2023Q4 \\
        Caixin PMI: Manufacturing & Production & Monthly & 2005Q3-2023Q4 \\
        PMI: Manufacturing Production & Production & Monthly & 2005Q1-2023Q4 \\
        Industrial Enterprises: Sales-output Ratio & Production & Monthly & 2012Q2-2023Q4 \\
        Industrial Enterprises: Export Delivery Value & Production & Monthly YoY & 1999Q1-2023Q4 \\
        Freight Traffic & Production & Monthly YoY & 2000Q1-2023Q4 \\
        Steel Output & Production & Monthly YoY & 1997Q1-2023Q4 \\
        Electricity Output & Production & Monthly YoY & 1997Q1-2023Q4 \\
        Non-Private Fixed Asset Investment & Investment & Monthly YoY & 1992Q1-2023Q4 \\
        Retail Sales of Consumer Goods & Consumption & Monthly YoY & 2003Q1-2023Q4 \\
        USA GDP & GDP International & Quarterly YoY & 1992Q1-2023Q4 \\
        Japan GDP & GDP International & Quarterly YoY & 1992Q1-2023Q4 \\
        Eurozone GDP & GDP International & Quarterly YoY & 1996Q1-2023Q4 \\
        Korea GDP & GDP International & Quarterly YoY & 1992Q1-2023Q4 \\
        \bottomrule
    \end{tabularx}
    \caption{Macroeconomic Data}
    \label{tab:macro-variable}
\end{table}

In the model presented in this paper, most macroeconomic variables used (except GDP) are in monthly frequency. In addition to the variables listed in Table \ref{tab:macro-variable}, we also tried to incorporate other macroeconomic and financial variables, such as M2, the Shanghai Composite Index, the national interbank offered rate, passenger car sales, industrial enterprise profits, etc. However, these variables are not included in the final models due to sampling periods or forecast performance. Prior to prediction, we conducted several data preprocessing steps. First, we convert volume variables to year-over-year growth values. Second, missing data are addressed through ARIMA imputation or forward filling.

To compare the forecasting results, this paper also utilizes forecasts from ``Longrun Expert Forecast" and ``Chief Economist Survey of Yicai Research Institute " (hereinafter ``Yicai Expert Forecast"). The ``Longrun Expert Forecast" is sourced from \cite{LiangFang2021}, encompassing 42 forecasts from Q1 2005 to Q4 2015.The "Longrun Expert Forecast" is initiated by the China Center for Economic Research at Peking University. At the invitation of the China Center for Economic Research, various economic research institutions have made forecasts for China’s macroeconomic indicators, such as GDP growth rate and CPI growth rate. \cite{LiangFang2021} primarily utilized the forecast results from the "Longrun Expert Forecast," including predictions from the China Center for Economic Research (CCER) and 11 other securities firms and research institutions. ``Yicai Expert Forecasts" are provided by Yicai Research Institute, comprising 30 forecasts from Q1 2014 to Q3 2023. ``Yicai Expert Forecasts" is initiated by Yicai Research Institute, inviting 18 chief economists to conduct surveys and forecasts on China's  macroeconomic indicators including GDP growth rate. Please note that data for Q3 2015 and Q4 2015-2023 are not reported in the ``Yicai Expert Forecasts".

\section{Methodology}
Table \ref{tab:model_list} lists all the models used in this paper, categorized into five groups. The first group (G1) comprises econometric models, including the autoregression (AR) model and the factor model (FM). The second group (G2) consists of machine learning models, including regularized linear regression models (LASSO, ridge regression), the kernel regression model (kernel ridge regression), and tree models (Random Forest, Gradient Boosting Tree, XGBoost). The third group (G3) includes combined models, which integrate the factor model with various machine learning models. The fourth group (G4) consists of mixed models, which are means and medians of group 2 and/or group 3 models. The final group (G5) is weighted models, which weight group 2 and group 3 models based on their performance.

\begin{table}
    \small
    \centering
    \begin{tabular}{c|c|cc}
    \toprule
    \textbf{\#}  & \textbf{Model Name} & \textbf{Model Details} & \textbf{Model Groups} \\
    \midrule
        1&AR&Autoregression&Econometric Model (base) - G1\\
        2&FM-AR-SE&Factor Model (FM)&Econometric Model - G1\\
        3&XGB-GBTREE&XGBoost&Machine Learning Model - G2\\
        4&XGB-GBLINEAR&XGBoost&Machine Learning Model - G2\\
        5&GBDT-AE&Gradient Boosting Decison Tree&Machine Learning Model - G2\\
        6&GBDT-HUBER&Gradient Boosting Decison Tree&Machine Learning Model - G2\\
        7&GBDT-SE&Gradient Boosting Decison Tree&Machine Learning Model - G2\\
        8&RF-AE&Random Forest (RF)&Machine Learning Model - G2\\
        9&RF-SE&Random Forest (RF)&Machine Learning Model - G2\\
        10&FM-XGB-GBLINEAR&FM+XGBoost&Combined Model - G3\\
        11&FM-XGB-GBTREE&FM+XGBoost&Combined Model - G3\\
        12&FM-GBDT-AE&FM+Gradient Boosting Decision Tree&Combined Model - G3\\
        13&FM-GBDT-HUBER&FM+Gradient Boosting Decision Tree&Combined Model - G3\\
        14&FM-GBDT-SE&FM+Gradient Boosting Decision Tree&Combined Model - G3\\
        15&FM-RF-AE&FM+RF&Combined Model - G3\\
        16&FM-RF-SE&FM+RF&Combined Model - G3\\
        17&FM-KRR-POLY&FM+Kernel Ridge Regression&Combined Model - G3\\
        18&FM-KRR-RBF&FM+Kernel Ridge Regression&Combined Model - G3\\
        19&FM-LASSO&FM+Lasso Regression&Combined Model - G3\\
        20&Median ML Models&Median of G1&Mixed Model -G4\\
        21&Mean ML Models&Mean of G1&Mixed Model -G4\\
        22&Median CC Models&Median of G2&Mixed Model -G4\\
        23&Mean CC Models&Mean of G2&Mixed Model -G4\\
        24&Median All Models&Median of G1 and G2&Mixed Model -G4\\
        25&Mean All Models&Mean of Group1 and 2&Mixed Model -G4\\
        26&RECIP4&Weighted by previous 4Q&Weighted Model - G5\\
        27&RECIP6&Weighted by previous 6Q&Weighted Model - G5\\
        28&RECIP8&Weighted by previous 8Q&Weighted Model - G5\\
        29&EXP0.5\_4&Previous 4Q exp 0.5&Weighted Model - G5\\
        30&EXP0.8\_4&Previous 4Q exp 0.8&Weighted Model - G5\\
        31&EXP0.9\_4&Previous 4Q exp 0.9&Weighted  Model - G5\\
        32&EXP1\_4&Previous 4Q exp 1&Weighted Model - G5\\
        33&EXP0.5\_6&Previous 6Q exp 0.5&Weighted Model - G5\\
        34&EXP0.8\_6&Previous 6Q exp 0.8&Weighted  Model - G5\\
        35&EXP0.9\_6&Previous 6Q exp 0.9&Weighted  Model - G5\\
        36&EXP1\_6&Previous 6Q exp 1&Weighted Model - G5\\
        37&EXP0.5\_8&Previous 8Q exp 0.5&Weighted Model - G5\\
        38&EXP0.8\_8&Previous 8Q exp 0.8&Weighted Model - G5\\
        39&EXP0.9\_8&Previous 8Q exp 0.9&Weighted Model - G5\\
        40&EXP1\_8&Previous 8Q exp 1&Weighted Model - G5\\
                 
    \bottomrule
    \end{tabular}
    \caption{List of Models}
    \label{tab:model_list}
\end{table}

\subsection{Single Model}
\subsubsection{Autoregressive Model}
The autoregressive (AR) model is a statistical method for analyzing time series data. The expected value of an explanatory variable is expressed as a linear combination of its own lagged historical data:

\begin{equation}
y_{t+h}=c+\phi(L)\ y_t+\varepsilon_{t+h},
\end{equation}
where $c$ is a constant term, $\Phi(L)$ is a $p$-order lag polynomial, $\epsilon_{t+h}$ is an error term, and $h$ represents the number of periods forecasted forward. As this method is a widely used traditional forecasting technique, the AR model serves as the benchmark for testing and comparing performance results in this paper.

\subsubsection{Factor Model}
Macroeconomic time series are defined by their data structure. The Bureau of Statistics of China publishes multidimensional economic and financial data, including monthly or quarterly figures on various variables related to production, consumption, employment, and finance. However, the years covered by these data are limited, resulting in a small number of observation samples that are difficult to expand. This abundance of time series, coupled with their relatively short lengths, can lead to estimation bias in forecasting.

The factor model can extract a relatively small number of latent factors from high-dimensional time series data, which are then used to forecast macroeconomic variables. The factor model's forecasting framework for macroeconomics is as follows:
\begin{equation}
y_{t+h}=c+\phi(L)\ y_t+\beta(L)\ F_t+\varepsilon_{t+h},
\end{equation}
\begin{equation}
X_t=\Lambda F_t+\eta_t,
\end{equation}
where $y$ is the macroeconomic variable to be predicted, $X$ is a high-dimensional time series in the economic and financial domain, and $F$ represents the latent factors. The prediction of $y$ in the forecasting model includes the lagged variables of $y$ itself, the latent factors and their lagged variables. Since a few latent factors can explain most of the variance in the macroeconomic series, the factor model effectively downscales the original macroeconomic data while fully utilizing the information within the original data.

\subsubsection{Machine Learning Model}
The machine learning models can be expressed in the following form:
\begin{equation}
y_{t+h}=f\left(Z_t\right)+\varepsilon_{t+h},
\end{equation}
\begin{equation}
Z_t=\left\{\left\{y_{t-j}\right\}_{j=p_0}^{p_y},\left\{X_{t-j}\right\}_{j=0}^{p_X}\right\},
\end{equation}
where $y$ is the macroeconomic variable to be forecasted, $Z$ is a characteristic variable constructed from the lagged variables of $y$ and the high-dimensional economic and financial time series $X$ and its lagged variables. The terms $p_y$ and $p_X$ represent the orders of the lagged variables of $y$ and $X$, respectively. The function $f$ is the forecasting function, $\varepsilon_{t+h}$ is the error term, and $h$ is the number of periods forecasted forward. The use of lagged variables is based on the assumption that macroeconomic variables depend on their past values.

In general, the essence of machine learning is to solve the following optimization problem:
\begin{equation}
\min_{f\in F} L(y_{t+h},f(Z_t))+R(f,\rho),
\end{equation}
where $y$ is the macroeconomic variable to be predicted, $Z$ is the high-dimensional time series feature variable used for prediction, $f$ is the prediction function, $F$ is the choice space of the prediction function, $L$ is the loss function, $R$ is the regularization term, and $\rho$ represents the model hyperparameters. The prediction function $f$ may be linear or nonlinear and parametric or nonparametric. The loss function can be a squared error loss, absolute error loss, or another type of loss function. The regularization term is designed in various ways to prevent overfitting. Hyperparameter selection includes information criteria (AIC or BIC) and the cross-validation method, with K-fold cross-validation primarily used in this paper.

The machine learning models used in this paper include Regularized Linear Regression (RLR) model, Kernel Ridge Regression (KRR) model, Random Forest (RF) model, Gradient Boosting Tree (GBDT) model, and Extreme Gradient Boosting (XGBoost) model. The main differences between different machine learning models come from the space of predictor functions, loss functions, regularization terms, hyperparameter selection, and the construction of feature variables.

\subsection{Multiple Models}

\subsubsection{Combined Model}
The combined model of machine learning and econometrics integrates econometric models (primarily the factor model) with machine learning models to leverage the strengths of both approaches. Initially, the factor model estimates the macroeconomic variable $X$ to obtain the factor. Subsequently, $y$, the factor, and their lagged variables are used as features for machine learning training. The mathematical expression of the coupled model is as follows:

\begin{equation}
y_{t+h}=f\left(Z_t\right)+\varepsilon_{t+h},
\end{equation}
\begin{equation}
X_t=\Lambda F_t+\eta_t,
\end{equation}
\begin{equation}
Z_t=\left\{\left\{y_{t-j}\right\}_{j=p_0}^{p_y},\left\{F_{t-j}\right\}_{j=0}^{P_F}\right\},
\end{equation}
where $X$ is the input macroeconomic variable, $F$ is the factor, $\Lambda$ is the factor loading, $f\left(\cdot\right)$ is the machine learning prediction function, \(Z\) is the feature set input to the machine learning model, and \(\varepsilon_{t+h}\) is the model error.

\subsubsection{Mixed Model}
The mixed model prediction methods generate new predictions by averaging or taking the median of results from multiple models. This paper employs various model combination methods, including machine learning models (group 2 models), combined models (group 3 models), and all machine learning-related models (group 2 and 3 models). The median of group prediction values or simple average of group prediction values (hereinafter referred to as the average prediction value) are employed. 

\subsubsection{Weighted Model}
The weighted model comprises each underlying machine learning model, with the weight of each model determined by its predictive performance function \(L\) (e.g., absolute error, squared error) over the past \(m\) quarters (where \(m = 4, 6, 8\)). Weights are assigned either by the inverse of the predictive performance function or by the exponential \(k\)th power of the performance function (where \(k = 0.5, 0.8, 0.9, 1\)). The mathematical expression is as follows:

\begin{equation}
{\hat{y}}_t=\sum_{j=1}^{N}w_{jt}{\hat{f}}_{jt}
\end{equation}

\begin{equation}
w_{jt}=\frac{1/\sum_{s=t-m}^{t-1}{L_{js}\left(y_s,{\hat{f}}_{js}\right\}}}{\sum_{k=1}^{M}{1/\sum_{s=t-m}^{t-1}{L_{ks}\left(y_s,{\hat{f}}_{ks}\right\}}}}
\end{equation}
or
\begin{equation}
w_{jt}=\frac{\exp{\left(-\beta\sum_{s=t-m}^{t-1}{L_{js}\left(y_s,{\hat{f}}_{js}\right)}\right)}}{\sum_{k=1}^{N}\exp{\left(-\beta\sum_{s=t-m}^{t-1}{L_{ks}\left(y_s,{\hat{f}}_{ks}\right)}\right)}}
\end{equation}

\subsection{Forecast Results Evaluation}

\subsubsection{Out-of-Sample Forecasts}
We evaluated the out-of-sample performance of the models through designed out-of-sample forecasting experiments. For forecasting China's quarterly GDP growth rate, we used the Expanding Window Method (EWM) for model training. Unlike the Rolling Window Method, which has a fixed window length, EWM increases the training data as the forecast time progresses. Given the relatively short time series of macroeconomic data, EWM facilitates the fuller utilization of historical macroeconomic information.

The specific forecasting steps are as follows. In the out-of-sample forecasting experiments, for forecasts at any given point, we assume that macroeconomic data in the "future" (post-forecasting point) are unknown to the model. The model is trained using historical macroeconomic data up to that point, with the current forecasting point data serving as the input vector to forecast the GDP growth rate for the current quarter. As more time points are forecasted, we compile the out-of-sample forecasting series of each model at various forecasting points. This out-of-sample forecast time series allows us to measure and evaluate the performance of different models. We forecast China's GDP growth rate from Q1 1996 to Q4 2023 using this out-of-sample forecasting method.

\begin{figure}
\centering
\includegraphics[width=0.8\textwidth]{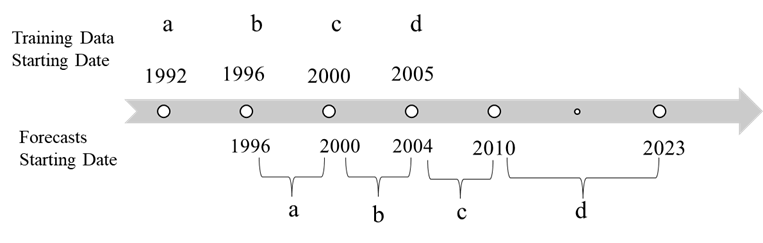}
\caption{Model Training and Forecasting Periods}
\label{fig:dates}
\end{figure}

The specific training data starting points and forecasting periods are illustrated in Figure \ref{fig:dates}. We divide the training data into four periods, labeled a to d, based on the earliest publication dates of various macroeconomic variables and their categories: before 1992 (a), before 1996 (b), before 2000 (c), and before 2005 (d). The years 1992, 1996, 2000, and 2005 mark the starting points for model training data in each period. Since machine learning models require historical data for training and estimation, the out-of-sample prediction intervals for each period begin after the training data start. For instance, in segment a, data from 1992 to 1995 are used to forecast China's GDP growth rate for Q1 1996; subsequent forecasts continue with an expanding window until the end of 1999. Similarly, segment b forecasts use macroeconomic data from 1996 to 1999 for training and forecast Q1 2000, continuing with an expanding window until the end of 2003. This methodology is applied to complete segment c forecasts for 2004-2009 and segment d forecasts for 2010-2023.

\subsection{Machine Learning Interpretability}

\subsubsection{Error Measurements}
For model metrics and evaluation, we use two metrics to assess out-of-sample performance: root mean square error (RMSE) and mean absolute error (MAE). In the robustness analysis, we measure the percentage reduction in error of the predictive model relative to the baseline model using out-of-sample squared or absolute errors. We define \({e}_{t,t,m} = y_t - \hat{y}_{t,h,m}\) as the prediction error of model \(m\) at time \(t\) for forecast horizon \(h\). The error measure is calculated as the average absolute error of the model at time \(h\). The specific error measures are calculated as follows:

The root mean squared error (RMSE) is:
\begin{equation}
RMSE_{h,m}=\left(\frac{1}{T}\sum_{t=1}^{T} e_{t,h,m}^2\right)^{1/2},
\end{equation}

The mean absolute error (MAE) is:
\begin{equation}
MAE_{h,m}=\frac{1}{T}\sum_{t=1}^{T}|e_{t,h,m}\mid.
\end{equation}

\subsubsection{Test Periods}
The performance of the models varies over time due to structural changes in the economy and exogenous shocks. This paper evaluates the out-of-sample forecasting ability of each model from Q1 1996 to Q1 2022 and compares their performance over time (see Figure\ref{fig:dates}). Section 5 specifically examines the models' out-of-sample forecasting ability during periods of high economic volatility, including the Asian Financial Crisis (1997-1998), the International Financial Crisis (2008-2010), and the COVID-19 pandemic (2020-2022).

In this paper, we adopt the Shapley Additive exPlanations (SHAP) framework to explain machine learning model predictions, as proposed by \cite{lundberg2017unified}. The SHAP approach is grounded in cooperative game theory, treating input variables as "contributors" to the machine learning model. By calculating the marginal contribution of each variable to the model prediction, we determine the Shapley value of each variable, revealing its importance to the macroeconomic forecast over time. This allows us to observe how different economic variables contribute to economic growth forecasts across various historical periods. Locally, the Shapley value enables analysis of each variable's contribution at each prediction time point. By aggregating local contributions over time, we can elucidate the functional relationship between economic variables and economic growth predictions, aiding economists and policymakers in better understanding and forecasting macroeconomic trends.

\section{Empirical Results}
In this section, we will discuss the prediction results of the model in detail. First, we will examine the overall performance of the machine learning model and compare it to that of the econometrics model and expert forecasting model. Next, we will analyze the main reasons for the differences in model results, particularly during periods of economic volatility and inflection points. Finally, we will evaluate the causes of model performance and assess the strengths and limitations of different machine learning models in macroeconomic forecasting.

\subsection{Forecasting Results}
\begin{figure}
\centering
\includegraphics[width=1\textwidth]{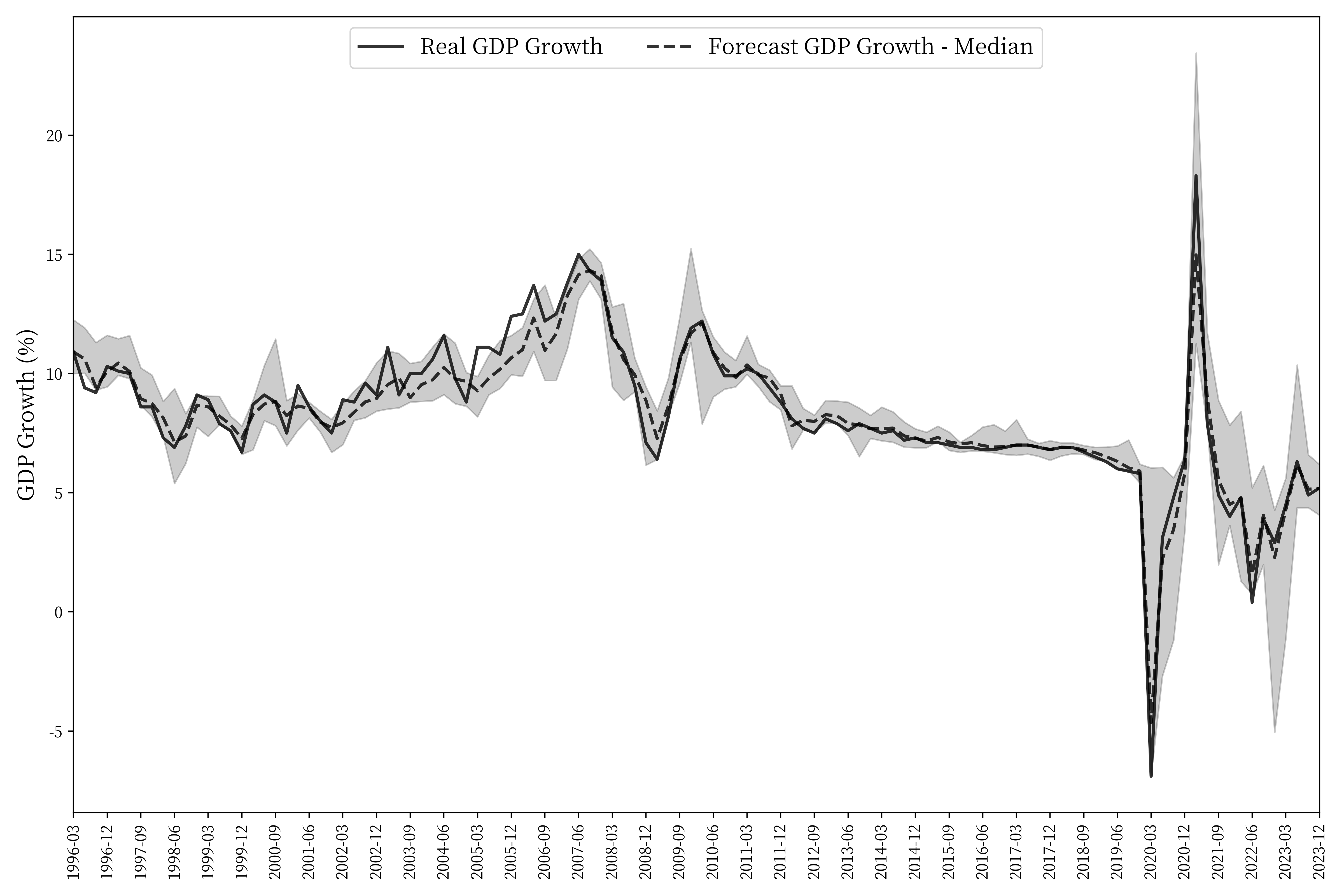}
\caption{Machine Learning Model Forecasts Median, Upper- and Lower-bounds: Quarterly Real GDP Growth}
\label{fig:full}
\end{figure}

Figure \ref{fig:full} illustrates the actual real GDP growth rate from Q1 1996 to Q4 2022 (solid line), the prediction range (gray area) for all single models (including both machine learning and combined models), and their median values (dashed line). Overall, the true value generally falls within the range of predicted values. Notably, the true value tends to converge towards the median prediction when there is no crisis.

In contrast, during periods of high economic volatility, such as the Asian financial crisis (1997-1998), the international financial crisis (2008-2010), and the COVID-19 pandemic (2020-2022), the model predictions still encompass the true value. However, the gap between the true value and the median prediction increases and the prediction range widens. For example, in Q1 2020, the GDP growth rate dropped to -6.9\% from 5.8\% in the previous quarter due to the COVID-19 pandemic. As shown in Figure \ref{fig:full}, the true value is closer to the predicted minimum, similar to Q1 2009 and Q2 2022.

In Table \ref{tab:modelcompare}, we compare the out-of-sample predictions of the machine learning models and combined models used in this paper with those of the econometric combination model used by \cite{LiangFang2021} and the ``Longrun Expert Forecasting" model. We selected the top three models in each category based on the smallest individual model prediction error (RMSE). The ``Longrun Expert Forecasts" data include 42 forecasts from Q3 2005 to Q4 2015, and the comparisons in Table \ref{tab:modelcompare} are limited to this period.    

\begin{table}[]
    \small
 \begin{tabularx}{\textwidth}{Xlc|Xlc}
\hline
\multicolumn{3}{c}{RMSE of models in this study}                         & \multicolumn{3}{c}{RMSE in \cite{LiangFang2021}}                    \\
\hline
\multirow{3}{=}{Machine Learning Model}                       & RF-AE         & 0.5845 & \multirow{3}{=}{Langrun Forecast}              & Best Model   & 0.6028 \\
                                                & XGB-GBTREE    & 0.5869 &                                                 & Second Best  & 0.6685 \\
                                                & RF-SE         & 0.5869 &                                                 & Third Best   & 0.6807 \\
\multicolumn{1}{l}{}                            &               &        & \multirow{3}{=}{Monthly Econometric Model}      & MIDAS        & 0.8107 \\
\multicolumn{1}{l}{}                            &               &        &                                                 & ECM-MIDAS    & 0.8151 \\
\multirow{3}{=}{Combined Model}                       & FM-RF-SE      & 0.7146 &                                                 & FM-VAR       & 0.8771 \\
                                                & FM-GBDT-AE    & 0.7262 & \multirow{3}{=}{Econometric Nonlinearity Model} & MS-VAR       & 0.843  \\
                                                & FM-GBDT-HUBER & 0.7434 &                                                 & TVTP         & 0.8561 \\
\multicolumn{1}{l}{}                            &               &        &                                                 & MS           & 0.8648 \\
\multicolumn{1}{l}{}                            &               &        & \multirow{3}{=}{Econometric Linear Model}       & ARMIA(1,1,1) & 0.8705 \\
\multirow{2}{=}{Econometric Model} & AR            & 1.1436 &                                                 & ARMIA(1,1,2) & 0.894  \\
                                                & FM-AR-SE      & 1.1932 &                                                 & ARMIA(2,1,1) & 0.9116\\
\hline
\end{tabularx}

    \caption{Model Performance Comparison}
    \label{tab:modelcompare}

\end{table}

Table \ref{tab:modelcompare} shows that, from Q3 2005 to Q4 2015, the machine learning models outperformed the ``Longrun Expert Forecast" and the econometric models in terms of forecast accuracy. The three machine learning models with the smallest errors had RMSEs below 0.6. The Longrun Expert Forecast and econometric models had RMSEs above 0.6 and 0.8, respectively. The combined models had a slightly higher error than the ``Longrun Expert Forecast", with an RMSE above 0.7, but still performed better than the traditional econometric model.

An important reason why machine learning models can make accurate predictions is their nonlinear characteristics (\citealp{coulombe2021can}), exemplified by models such as random forest and kernel ridge regression. Econometric models typically use linear functions to predict the dependent variable. However, the functional relationship between independent and dependent variables is often not strictly linear in reality. Machine learning models with nonlinear characteristics can therefore improve prediction accuracy. This advantage is confirmed by comparing the results of \cite{LiangFang2021} econometric model. Table 3 shows that the errors of nonlinear econometric models (MS-VAR, TVTP, MS) are smaller than those of linear econometric models (ARIMA(1,1,1), ARIMA(1,1,2), ARIMA(2,1,1)). The models associated with the random forest model perform relatively better during this period. However, the performance of machine learning models can fluctuate across different periods, and random forest-related models may not always be the best performers, as seen in results from other periods.

In Table \ref{tab:modelcompare2}, we compare the out-of-sample forecasts of three types of models (machine learning models, combined models, and all machine learning-related models) with the econometric combination model used by \cite{LiangFang2021} and the combination model from the ``Longrun Expert Forecast". The comparison period spans from Q3 2005 to Q4 2015.

\begin{table}[]
    \small
\begin{tabularx}{\textwidth}{X|XX}
\hline
\multicolumn{3}{c}{Composite RMSE (\citealp{LiangFang2021})}                                                                          \\
\hline

                                                                        & Mean value                           & 0.6528               \\
\multirow{-2}{=}{Langrun Forecast}                                     & Median value                         & 0.6528 \\
                                                                        & Mean value                           & 0.8701 \\

\multirow{-2}{=}{Econometric Model}                                   & Median value                         & 0.8701               \\

\hline
\multicolumn{3}{c}{Composite RMSE (this study)}                                                                                     \\
\hline

                                                                        & Mean value                           & 0.5702               \\
\multirow{-2}{=}{Machine Learning Model (G2)}                         & Median value                         & 0.5197               \\
                                                                        & Mean value                           & 0.7599               \\

\multirow{-2}{=}{Combined Model (G3)}                                 & Median value                         & 0.7745               \\
                                                                        & Mean value                           & 0.6464               \\

\multirow{-2}{=}{Machine Learning Model \& Combined Model (G2 \& G3)} & Median value                         & 0.6282               \\
                                                & RECIP\_4     & 0.6144               \\
                                                & RECIP\_6     & 0.6285               \\
                                                & RECIP\_8     & 0.6364               \\
                                                & EXP\_-0.5\_4 & 0.6487               \\
                                                & EXP\_-0.8\_4 & 0.6425               \\
                                                & EXP\_-0.9\_4 & 0.6406               \\
                                                & EXP\_-1\_4   & 0.6387               \\
                                                & EXP\_-0.5\_6 & 0.6535               \\
                                                & EXP\_-0.8\_6 & 0.6498               \\
                                                & EXP\_-0.9\_6 & 0.6486               \\
                                                & EXP\_-1\_6   & 0.6474               \\
                                                & EXP\_-0.5\_8 & 0.6548               \\
                                                & EXP\_-0.8\_8 & 0.6517               \\
                                                & EXP\_-0.9\_8 & 0.6507               \\

\multirow{-15}{=}{Weighted Model (G4)}     & EXP\_-1\_8                           & 0.6498 \\             
\hline
\end{tabularx}

    \caption{Model Performance Comparison2}
    \label{tab:modelcompare2}
\end{table}

The results in Table \ref{tab:modelcompare2} are close to those in Table \ref{tab:modelcompare}. The machine learning model performs best, followed by all machine learning related models. The ``Langrun Expert Forecast" outperforms the combined and econometric model forecasts.

In order to compare the prediction results of the recent models, Table \ref{tab:modelcompare3} shows the results of the models after 2014, and the out-of-sample prediction results with the ``Yicai Expert Forecast". The ``Yicai Expert Forecast" data has a total of 30 forecasts from 2014 to 2023. We exclude several data points that are not reported in the ``Yicai Expert Forecasts" from the comparisons, such as the Q3 2015 and Q4 2015-2023.

\begin{table}[]
    \scriptsize 
\begin{tabularx}{\textwidth}{XX|cc|cc|cc|cc}
\hline
\multirow{2}{*}{Groups} & \multirow{2}{*}{\textbf{Model Names}} & \multicolumn{2}{c|}{2014-2023} & \multicolumn{2}{c|}{2014-2019} & \multicolumn{2}{c|}{2020-2022} & \multicolumn{2}{c}{2023} \\

\multicolumn{2}{c|}{}                                                    & RMSE          & MAE           & RMSE          & MAE           & RMSE          & MAE           & RMSE        & MAE        \\
\hline
\multirow{2}{=}{Machine Learning Model}                               & Mean value          & 1.17          & 0.68          & 0.22          & 0.18          & 2.16          & 1.81          & 3.59        & 3.51       \\
                                                  & Median value        & 1.23          & 0.55          & 0.19          & 0.15          & 2.35          & 1.69          & 3.79        & 3.71       \\
\hline
\multirow{2}{=}{Combined Model}                               & Mean value          & 1.05          & 0.57          & 0.15          & 0.11          & 1.91          & 1.65          & 0.13        & 0.11       \\
                                                  & Median value        & 0.85          & 0.45          & 0.13          & 0.09          & 1.53          & 1.29          & 0.14        & 0.12       \\
\hline
\multirow{2}{=}{Machine Learning \& Combined Model}                        & Mean value          & 1.01          & 0.54          & 0.15          & 0.12          & 1.81          & 1.45          & 0.41        & 0.37       \\
                                                  & Median value        & 0.86          & 0.45          & 0.14          & 0.11          & 1.55          & 1.21          & 0.20        & 0.20       \\
\hline
\multirow{15}{=}{Weighted Model}        & RECIP\_4            & 0.82          & 0.47          & 0.14          & 0.11          & 1.42          & 1.18          & 0.51        & 0.47       \\
                                                  & RECIP\_6            & 0.84          & 0.49          & 0.14          & 0.11          & 1.45          & 1.24          & 0.55        & 0.44       \\
                                                  & RECIP\_8            & 0.84          & 0.49          & 0.15          & 0.12          & 1.51          & 1.28          & 0.41        & 0.34       \\
                                                  & EXP\_0.5\_4         & 1.02          & 0.56          & 0.17          & 0.14          & 1.83          & 1.47          & 0.46        & 0.37       \\
                                                  & EXP\_0.8\_4         & 0.98          & 0.54          & 0.17          & 0.14          & 1.75          & 1.39          & 0.49        & 0.39       \\
                                                  & EXP\_0.9\_4         & 0.97          & 0.53          & 0.17          & 0.14          & 1.73          & 1.37          & 0.50        & 0.41       \\
                                                  & EXP\_1\_4           & 0.96          & 0.53          & 0.17          & 0.13          & 1.71          & 1.35          & 0.51        & 0.42       \\
                                                  & EXP\_0.5\_6         & 1.05          & 0.58          & 0.17          & 0.14          & 1.89          & 1.53          & 0.45        & 0.39       \\
                                                  & EXP\_0.8\_6         & 1.02          & 0.57          & 0.17          & 0.14          & 1.82          & 1.48          & 0.48        & 0.40       \\
                                                  & EXP\_0.9\_6         & 1.01          & 0.56          & 0.17          & 0.14          & 1.80          & 1.47          & 0.49        & 0.41       \\
                                                  & EXP\_1\_6           & 1.00          & 0.56          & 0.17          & 0.14          & 1.78          & 1.46          & 0.50        & 0.41       \\
                                                  & EXP\_0.5\_8         & 1.07          & 0.58          & 0.17          & 0.14          & 1.92          & 1.55          & 0.41        & 0.36       \\
                                                  & EXP\_0.8\_8         & 1.04          & 0.57          & 0.17          & 0.14          & 1.86          & 1.52          & 0.41        & 0.36       \\
                                                  & EXP\_0.9\_8         & 1.03          & 0.57          & 0.17          & 0.14          & 1.84          & 1.50          & 0.41        & 0.36       \\
                                                  & EXP\_1\_8           & 1.02          & 0.57          & 0.17          & 0.14          & 1.83          & 1.49          & 0.41        & 0.36       \\
\hline
\multirow{2}{*}{Yicai Expert Forecast}                     & Mean value          & 0.35          & 0.29          & 0.17          & 0.15          & 0.51          & 0.45          & 0.55        & 0.54       \\
                                                  & Median value        & 0.32          & 0.25          & 0.17          & 0.15          & 0.48          & 0.39          & 0.50        & 0.43  \\

\hline
\end{tabularx}
    \caption{Comparison of Derivative Composite Model Performance (2014-2023)}
    \label{tab:modelcompare3}   

\end{table}

Table \ref{tab:modelcompare3} shows that from 2014 to 2019, when the economy was relatively stable, both  machine learning models and combined models outperformed the ``Yicai Expert Forecast". However, during 2014 to 2024, “Yicai Expert Forecast” is more accurate. By comparing the results in different economic periods, we find that “Yicai Expert Forecast” is particularly more accurate during the COVID-19 pandemic period after 2020, with significantly lower prediction errors compared to all other types of models. In the next section, we will further discuss the model prediction performance and its causes during periods of high economic fluctuations or at economic inflection points.

\subsection{Model Comparisons}

\subsubsection{Forecast Performance during Crisis}

\begin{figure}
\centering
\includegraphics[width=0.8\textwidth]{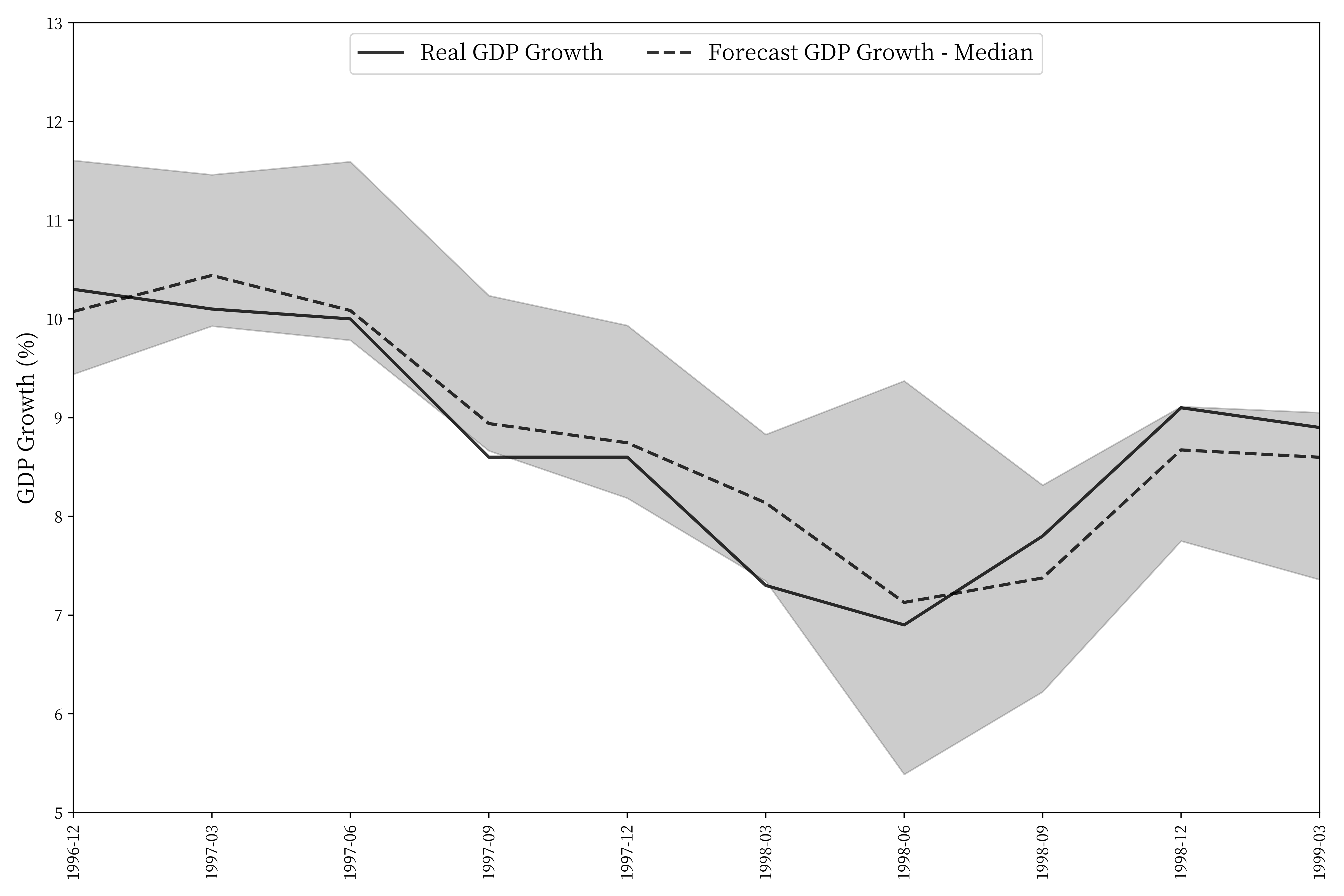}
\caption{Asian Financial Crisis, 1997-1998}
\label{fig:1997-1998}
\end{figure}

\begin{figure}
\centering
\includegraphics[width=0.8\textwidth]{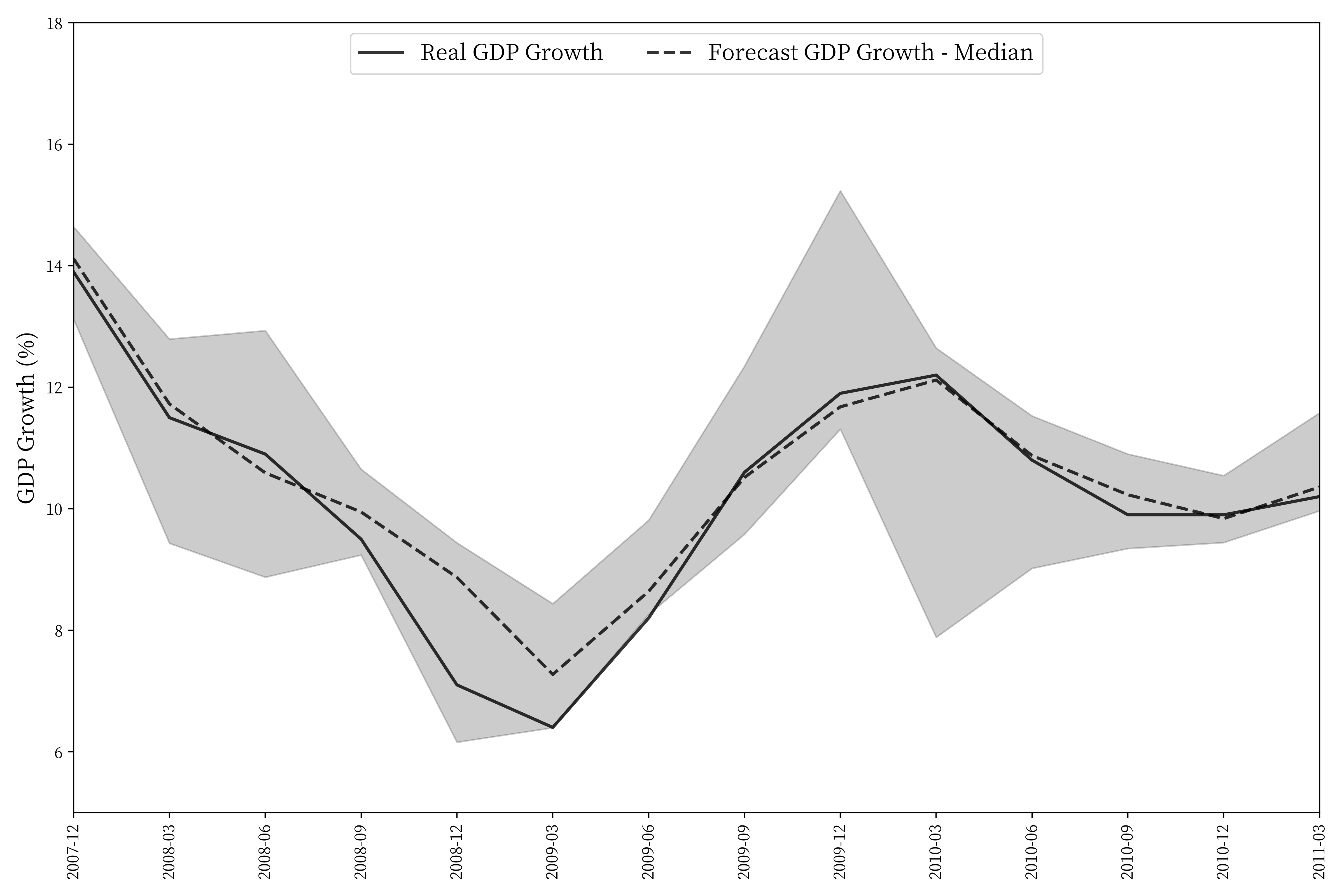}
\caption{Global Financial Crisis, 2008-2010}
\label{fig:2008-2010}
\end{figure}

\begin{figure}
\centering
\includegraphics[width=0.8\textwidth]{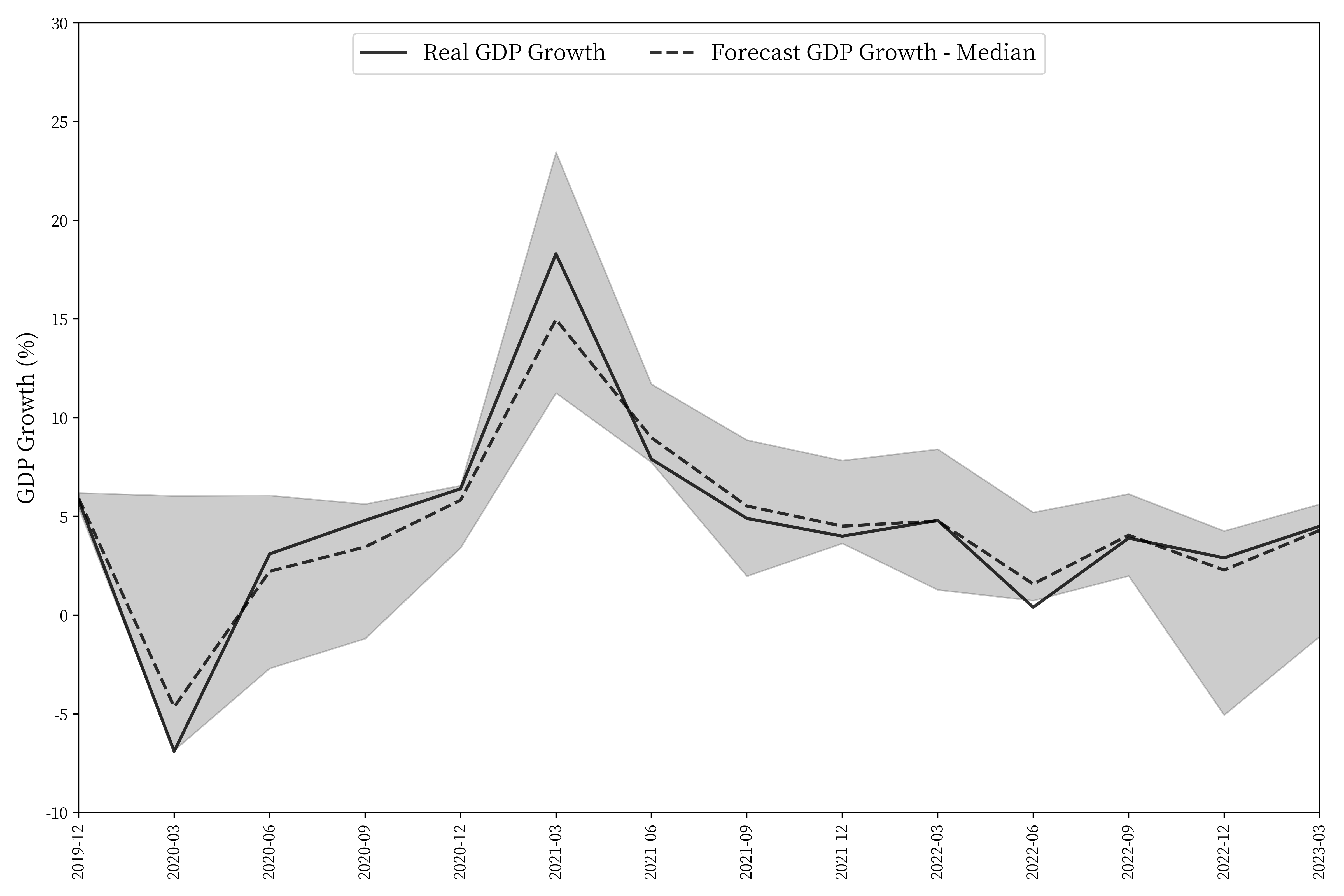}
\caption{Covid-19 Epidemic Period, 2020-2022}
\label{fig:2020-2022}
\end{figure}

Figures  \ref{fig:1997-1998}, \ref{fig:2008-2010}, and \ref{fig:2020-2022} plot the time series of true values and out-of-sample forecasts for three periods of economic volatility: the 1997-1998 Asian Financial Crisis, the 2008-2010 Global Financial Crisis, and the 2020-2022 COVID-19 pandemic, respectively. We selected the performance of the median predictions of machine-learning related models (G2 and G3 models) to compare with the actual real GDP growth rates.

As shown in Figures \ref{fig:1997-1998} through \ref{fig:2020-2022}, machine learning models accurately predict the economic inflection points during these three periods, demonstrating a high degree of accuracy in determining the direction of change. Comparing the median prediction values of the models with the true values reveals that, although economic fluctuations during the 2008-2010 Global Financial Crisis were greater than those during the 1997-1998 Asian Financial Crisis, the prediction accuracy of the machine learning models did not decrease significantly. For instance, the median RMSE of the pure machine learning model was 0.53 during the 1997-1998 period and 0.36 during the 2008-2010 period. This improved accuracy can be attributed to the richer and more variable data available for training the machine learning models during the 2008-2010 period (see Table \ref{tab:macro-variable}).

For the 2020-2022 period during the COVID-19 pandemic, the prediction accuracy of the machine learning model, particularly for predicting inflection points, is lower than during the 2008-2010 period, despite the training data containing more variables. The primary reason for this is the highly heterogeneous nature of the economic shock caused by the pandemic compared to the financial crisis. The economic shutdowns due to city lockdowns had a completely different mechanism from the impact of the global financial crisis on China's export demand. Even with the inclusion of previous economic fluctuation data in the training set, the prediction accuracy during the COVID-19 period did not substantially improve.

In the literature review, we mentioned that \cite{ho2023forecasting} summarized the macroeconomic forecasting methodology during the COVID-19 pandemic. Ho found that, due to the unpredictability of events such as the pandemic and the unique nature of its economic shocks, it is particularly important to include real-time information (e.g., data on the spread of the epidemic) in forecasts. This information must be used with a solid understanding of economics and the core drivers behind economic fluctuations. This is where expert forecasting has a significant advantage over machine learning models (see Table \ref{tab:modelcompare3}).

Second, economic volatility during the COVID-19 pandemic was significantly greater than during the previous two financial crises. Quarterly GDP growth turned negative for the first time, with the minimum (maximum) values of the true GDP being much lower (higher) than the minimum (maximum) values of the training data. In such cases, tree-based machine learning models, such as the gradient boosted tree model (GBDT) and the random forest model (RF), find it more challenging to accurately predict inflection point values. Conversely, kernel ridge regression (KRR) does not face this limitation (\citep{coulombe2021can}). As a result, the ridge regression (KRR) model demonstrated higher prediction accuracy than other machine learning models during the 2020-2022 COVID-19 pandemic period.

A comparison of recent prediction results of machine learning models (Q1 2014 - Q4 2023) is presented in Table \ref{tab:modelcompare4}. It shows that the two machine learning models associated with ridge regression (KRR)—FM-KRR-POLY and FM-KRR-RBF—have significantly higher prediction accuracies than other machine learning models during the 2020-2022 COVID-19 pandemic period. The RMSE and MAE of other models are several times higher than those of these two models. Additionally, from Q1 2014 to Q4 2019, FM-KRR-POLY and FM-KRR-RBF also exhibit better prediction results compared to other machine learning models. This implies that the ridge regression models predict inflection points more accurately than tree-based models, gradient boosted trees, and random forests when the fluctuations in the actual inflection point data are much larger than those in the model training data.

\begin{table}[]
    \scriptsize 
\begin{tabularx}{\textwidth}{Xl|cc|cc|cc|cc}
\hline
\multirow{2}{*}{Groups} & \multirow{2}{*}{\textbf{Model Names}} & \multicolumn{2}{c|}{2014-2023} & \multicolumn{2}{c|}{2014-2019} & \multicolumn{2}{c|}{2020-2022} & \multicolumn{2}{c}{2023} \\

\multicolumn{2}{c|}{}                                                    & RMSE          & MAE           & RMSE          & MAE           & RMSE          & MAE           & RMSE        & MAE        \\
\hline
\multirow{7}{=}{Machine Learning Model} & GBDT-AE         & 2.37            & 0.89            & 0.22            & 0.17            & 4.28            & 2.38            & 0.95        & 0.70       \\
                                             & GBDT-HUBER      & 1.02            & 0.52            & 0.18            & 0.15            & 1.74            & 1.12            & 1.05        & 0.95       \\
                                             & GBDT-SE         & 1.37            & 0.68            & 0.18            & 0.14            & 2.39            & 1.71            & 1.25        & 0.84       \\
                                             & RF-AE           & 1.45            & 0.75            & 0.19            & 0.14            & 2.28            & 1.69            & 2.30        & 1.51       \\
                                             & RF-SE           & 2.02            & 1.10            & 0.26            & 0.22            & 3.38            & 2.55            & 2.45        & 2.03       \\
                                             & XGB-GBLINEAR    & 1.51            & 0.98            & 0.63            & 0.51            & 2.57            & 2.03            & 0.72        & 0.66       \\
                                             & XGB-GBTREE      & 1.41            & 0.57            & 0.17            & 0.14            & 1.99            & 1.05            & 2.82        & 1.67       \\
 \hline
\multirow{10}{=}{Combined Model}        & FM-GBDT-AE      & 1.80            & 0.80            & 0.25            & 0.16            & 3.26            & 2.28            & 0.26        & 0.20       \\
                                             & FM-GBDT-HUBER   & 2.27            & 0.93            & 0.17            & 0.13            & 4.13            & 2.66            & 0.62        & 0.52       \\
                                             & FM-GBDT-SE      & 1.59            & 0.85            & 0.18            & 0.14            & 2.84            & 2.29            & 0.85        & 0.80       \\
                                             & FM-KRR-POLY     & 0.57            & 0.37            & 0.20            & 0.14            & 0.92            & 0.78            & 0.62        & 0.48       \\
                                             & FM-KRR-RBF      & 0.57            & 0.37            & 0.21            & 0.15            & 0.92            & 0.79            & 0.63        & 0.49       \\
                                             & FM-LASSO        & 0.94            & 0.49            & 0.15            & 0.11            & 1.68            & 1.25            & 0.53        & 0.47       \\
                                             & FM-RF-AE        & 1.59            & 0.80            & 0.21            & 0.16            & 2.86            & 2.19            & 0.55        & 0.47       \\
                                             & FM-RF-SE        & 1.10            & 0.64            & 0.25            & 0.19            & 1.96            & 1.67            & 0.33        & 0.26       \\
                                             & FM-XGB-GBLINEAR & 1.66            & 0.95            & 0.35            & 0.28            & 2.94            & 2.39            & 1.00        & 0.69       \\
                                             & FM-XGB-GBTREE   & 1.39            & 0.74            & 0.23            & 0.20            & 2.45            & 1.82            & 1.03        & 0.71       \\
\hline
\multirow{15}{=}{Weighted Model}   & RECIP\_4        & 0.77            & 0.46            & 0.14            & 0.12            & 1.38            & 1.17            & 0.44        & 0.36       \\
                                             & RECIP\_6        & 0.78            & 0.47            & 0.15            & 0.12            & 1.39            & 1.22            & 0.48        & 0.35       \\
                                             & RECIP\_8        & 0.78            & 0.46            & 0.15            & 0.12            & 1.39            & 1.20            & 0.36        & 0.26       \\
                                             & EXP\_0.5\_4     & 0.94            & 0.52            & 0.17            & 0.14            & 1.68            & 1.36            & 0.40        & 0.28       \\
                                             & EXP\_0.8\_4     & 0.90            & 0.50            & 0.17            & 0.14            & 1.61            & 1.31            & 0.42        & 0.31       \\
                                             & EXP\_0.9\_4     & 0.89            & 0.50            & 0.17            & 0.14            & 1.60            & 1.29            & 0.43        & 0.32       \\
                                             & EXP\_1\_4       & 0.89            & 0.50            & 0.17            & 0.14            & 1.58            & 1.28            & 0.44        & 0.33       \\
                                             & EXP\_0.5\_6     & 0.96            & 0.54            & 0.17            & 0.14            & 1.72            & 1.41            & 0.39        & 0.30       \\
                                             & EXP\_0.8\_6     & 0.93            & 0.53            & 0.17            & 0.14            & 1.67            & 1.38            & 0.42        & 0.31       \\
                                             & EXP\_0.9\_6     & 0.92            & 0.52            & 0.17            & 0.14            & 1.65            & 1.37            & 0.43        & 0.32       \\
                                             & EXP\_1\_6       & 0.92            & 0.52            & 0.17            & 0.14            & 1.64            & 1.36            & 0.43        & 0.32       \\
                                             & EXP\_0.5\_8     & 0.97            & 0.54            & 0.17            & 0.14            & 1.74            & 1.43            & 0.35        & 0.27       \\
                                             & EXP\_0.8\_8     & 0.94            & 0.53            & 0.17            & 0.14            & 1.69            & 1.39            & 0.35        & 0.27       \\
                                             & EXP\_0.9\_8     & 0.94            & 0.53            & 0.17            & 0.14            & 1.68            & 1.38            & 0.35        & 0.27       \\
                                             & EXP\_1\_8       & 0.93            & 0.52            & 0.17            & 0.14            & 1.67            & 1.37            & 0.35        & 0.27  \\
\hline
 \end{tabularx}                                            \\
    \caption{Comparison of the Machine Learning Models' Performance (2014-2023)}
    \label{tab:modelcompare4}  

\end{table}

Table  \ref{tab:modelcompare5} lists the machine learning model prediction results by different time periods for 2005-2015. It shows that the XGB-GBTREE model achieves the best prediction results among individual models during the 2008-2010. The median machine learning model has relativly high prediction accuracy both during the overall 2005-2015 period and during the 2008-2010 Global Financial Crisis (GFC) period. Conversely, the two ridge regression (KRR) models (FM-KRR-POLY and FM-KRR-RBF) exhibited higher errors during the 2008-2010 period. This indicates that the ridge regression models did not outperform other machine learning models when the economic growth is relatively stable.

\begin{table}[]
  \scriptsize 
\begin{tabularx}{\textwidth}{Xc|cc|cc|cc}
\hline
\multicolumn{2}{c|}{Sampling Periods}                                                              & \multicolumn{2}{|c|}{2005Q3-2015Q4}                           & \multicolumn{2}{|c|}{2008-2010}                               & \multicolumn{2}{|c}{exl.2008-2010}                           \\
\hline
\multicolumn{2}{c|}{Single Model Comparison}                                           & RMSE                         & MAE                          & RMSE                         & MAE                          & RMSE                         & MAE                          \\
\hline
                                                & RF-AE                               & 0.58                         & 0.44                         & 0.62                         & 0.51                         & 0.57                         & 0.41                         \\
                                                & \cellcolor[HTML]{D9D9D9}XGB-GBTREE  & \cellcolor[HTML]{D9D9D9}0.59 & \cellcolor[HTML]{D9D9D9}0.43 & \cellcolor[HTML]{D9D9D9}0.39 & \cellcolor[HTML]{D9D9D9}0.27 & \cellcolor[HTML]{D9D9D9}0.65 & \cellcolor[HTML]{D9D9D9}0.50 \\
\multirow{-3}{=}{Machine Learning Model (top three)}         & RF-SE                               & 0.59                         & 0.44                         & 0.57                         & 0.45                         & 0.59                         & 0.43                         \\
                                                & FM-RF-SE                            & 0.71                         & 0.47                         & 0.96                         & 0.75                         & 0.59                         & 0.36                         \\
                                                & FM-GBDT-AE                          & 0.73                         & 0.50                         & 0.82                         & 0.65                         & 0.69                         & 0.44                         \\
                                                & FM-GBDT-HUBER                       & 0.74                         & 0.51                         & 0.77                         & 0.58                         & 0.73                         & 0.48                         \\
                                                & \cellcolor[HTML]{D9D9D9}FM-KRR-POLY & \cellcolor[HTML]{D9D9D9}1.24 & \cellcolor[HTML]{D9D9D9}0.85 & \cellcolor[HTML]{D9D9D9}1.61 & \cellcolor[HTML]{D9D9D9}1.18 & \cellcolor[HTML]{D9D9D9}1.04 & \cellcolor[HTML]{D9D9D9}0.71 \\
\multirow{-5}{=}{Combined Model (top three and KRR)} & \cellcolor[HTML]{D9D9D9}FM-KRR-RBF  & \cellcolor[HTML]{D9D9D9}1.25 & \cellcolor[HTML]{D9D9D9}0.86 & \cellcolor[HTML]{D9D9D9}1.62 & \cellcolor[HTML]{D9D9D9}1.18 & \cellcolor[HTML]{D9D9D9}1.06 & \cellcolor[HTML]{D9D9D9}0.72 \\
                                                & AR                                  & 1.14                         & 0.88                         & 1.43                         & 1.17                         & 1.01                         & 0.77                         \\
\multirow{-2}{=}{Econometric Model}             & FM-AR-SE                            & 1.19                         & 0.82                         & 1.68                         & 1.33                         & 0.91                         & 0.60                         \\
\hline
\multicolumn{2}{c|}{Multiple Model Comparison}                                        & RMSE                         & MAE                          & RMSE                         & MAE                          & RMSE                         & MAE                          \\
\hline
                                                & Mean     & 0.57                         & 0.41                         & 0.39                         & 0.32                         & 0.64                         & 0.45                         \\
\multirow{-2}{=}{Machine Learning Mixed Model}                      & Median    & 0.52                         & 0.37                         & 0.36                         & 0.29                         & 0.58                         & 0.41                         \\
                                                & Mean     & 0.76                         & 0.54                         & 0.80                         & 0.60                         & 0.76                         & 0.53                         \\
\multirow{-2}{=}{Combined Mixed Model}                      & Median    & 0.77                         & 0.50                         & 0.73                         & 0.49                         & 0.81                         & 0.52                         \\
                                                &Mean     & 0.65                         & 0.46                         & 0.57                         & 0.44                         & 0.69                         & 0.48                         \\
\multirow{-2}{=}{Machine Learning \& Combined Mixed Model}              & Median   & 0.63                         & 0.43                         & 0.62                         & 0.41                         & 0.64                         & 0.44                         \\
                                                & RECIP\_4                            & 0.61                         & 0.43                         & 0.48                         & 0.36                         & 0.66                         & 0.46                         \\
                                                & RECIP\_6                            & 0.63                         & 0.44                         & 0.53                         & 0.41                         & 0.66                         & 0.45                         \\
                                                & RECIP\_8                            & 0.64                         & 0.45                         & 0.55                         & 0.42                         & 0.67                         & 0.46                         \\
                                                & EXP\_0.5\_4                         & 0.65                         & 0.46                         & 0.59                         & 0.45                         & 0.67                         & 0.46                         \\
                                                & EXP\_0.8\_4                         & 0.64                         & 0.45                         & 0.58                         & 0.44                         & 0.67                         & 0.46                         \\
                                                & EXP\_0.9\_4                         & 0.64                         & 0.45                         & 0.58                         & 0.43                         & 0.66                         & 0.46                         \\
                                                & EXP\_1\_4                           & 0.64                         & 0.45                         & 0.57                         & 0.43                         & 0.66                         & 0.46                         \\
                                                & EXP\_0.5\_6                         & 0.65                         & 0.46                         & 0.60                         & 0.46                         & 0.67                         & 0.46                         \\
                                                & EXP\_0.8\_6                         & 0.65                         & 0.46                         & 0.59                         & 0.46                         & 0.67                         & 0.46                         \\
                                                & EXP\_0.9\_6                         & 0.65                         & 0.46                         & 0.59                         & 0.46                         & 0.67                         & 0.46                         \\
                                                & EXP\_1\_6                           & 0.65                         & 0.46                         & 0.59                         & 0.46                         & 0.67                         & 0.46                         \\
                                                & EXP\_0.5\_8                         & 0.65                         & 0.47                         & 0.60                         & 0.46                         & 0.68                         & 0.47                         \\
                                                & EXP\_0.8\_8                         & 0.65                         & 0.46                         & 0.59                         & 0.46                         & 0.67                         & 0.46                         \\
                                                & EXP\_0.9\_8                         & 0.65                         & 0.46                         & 0.59                         & 0.46                         & 0.67                         & 0.46                         \\
\multirow{-15}{=}{Weighted Composite Model}     & EXP\_1\_8                           & 0.65                         & 0.46                         & 0.59                         & 0.46                         & 0.67                         & 0.46 \\
\hline
\end{tabularx}
    \caption{Comparison of Model Performance by Periods (2005Q3-2015Q4)}
    \label{tab:modelcompare5}  
\end{table}

Additionally, the prediction performance of the multiple models (mixed model or weighted models) is more stable. Although prediction errors increase during periods of higher economic volatility, the increase is much lower than that of single models. Besides, during periods when the economic growth is stable, multiple models' prediction accuracy surpasses that of the expert forecasts (see Tables \ref{tab:modelcompare2} and \ref{tab:modelcompare3})). Therefore, outcomes of multiple models hold high reference value.

\subsubsection{Inclusive Testing}
As shown in Table \ref{tab:modelcompare3}, ``Yicai Expert Forecast" from 2014 to 2023 has an advantage over the machine learning models. One important reason for this could be that experts would utilizes more real-time information than the machine learning model. 

To test this hypothesis, we use the inclusive test as follows:
\begin{equation}
e_{t,i}=\left(1-\lambda\right)\left(e_{t,i}-e_{t,j}\right)+\varepsilon,
\end{equation}
where $e_{t,i}$ and $e_{t,j}$ denote the errors of the ``Yicai Expert Forecast" and the mixed model predictions, respectively. If $\left(1-\lambda\right)=1$, it indicates that the information in the ``Yicai Expert Forecast" is less than that covered by the machine learning model prediction. Conversely, if $\left(1-\lambda\right)=0$, it means that the ``Yicai Expert Forecast" encompasses the information predicted by the machine learning model. We perform OLS regression on the above equation to obtain the estimated coefficients of $1-\lambda$ to determine the information inclusion relationship between ``Yicai Expert Prediction" and mixed model predictions.

Table \ref{tab:modelcompare6} reports the estimated coefficients of $1-\lambda$ and the corresponding t-statistics. The test results indicate that the estimated values of $1-\lambda$ are all around 0. The t-statistics are not significant in most tests (the p-value is greater than 0.05), and the original hypothesis of $\left(1-\lambda\right)=0$ cannot be rejected. This suggests that the ``Yicai Expert Forecast" contains the information predicted by the mixed model. Additionally, during periods of high economic volatility, expert forecasts tend to be more accurate in predicting inflection points. In other words, experts could leverage their knowledge of economics to make rapid judgments about economic shocks or public policy releases, performing better than machine learning models during these inflection points. Therefore, during periods of high economic volatility, machine learning macro forecasting models need to incorporate additional variables, such as high-frequency public health information, freight, traffic, news or economic policies changes.

\begin{table}[]
\small
\begin{tabular}{c|c|cc|cc|cc}
\hline
\multicolumn{2}{c}{\multirow{2}{*}{1-$\lambda$}}    & \multicolumn{2}{|c|}{ML Model} & \multicolumn{2}{c|}{Combined Model} & \multicolumn{2}{c}{ML and Combined Model} \\

\multicolumn{2}{c|}{}                        & Mean           & Median        & Mean                   & Median                & Mean            & Median         \\
\hline
\multirow{4}{*}{Yicai Expert} & Mean        & 0.037          & 0.048         & 0.020                  & 0.039                 & 0.083           & 0.160*         \\
                              & t-statistics & (0.03)         & (0.03)        & (0.04)                 & (0.04)                & (0.06)          & (0.06)         \\
                              & Median      & 0.013          & 0.012         & 0.002                  & 0.022                 & -0.076          & -0.011         \\
                              & t-statistics & (0.03)         & (0.03)        & (0.03)                 & (0.04)                & (0.07)          & (0.08)        \\
\hline
\end{tabular}

    \caption{Inclusive Test of GDP Projections (2014-2023)}
    \label{tab:modelcompare6} 
\end{table}

\section{Interpretability Analysis}

\subsection{Global Perspective}
Shapley values reflect the marginal impact of a variable on the final predicted value for the entire sample. Shapley value analysis is currently the most general and rigorous approach to addressing the issues of machine learning macro prediction interpretability and model communication. By using Shapley values, the importance of each variable can be assessed globally, explaining the drivers of model predictions and performance.

Tables \ref{tab:shapley_ML_full} and \ref{tab:sharpleyFMfull} present the Shapley values for each model from 2006 to 2023, with the gray areas indicating the top five most important variables in each model. The results show that, due to the downscaling of variables in the factor model measures, the key driving variables differ from those in other models. For the factor model and its associated combined models (see Table \ref{tab:sharpleyFMfull} ), the more important variables are: area of new housing construction, area of sales of commercial properties, amount of imports, amount of exports, and amount of completed non-private fixed asset investment. However, for the non-factor model (see Table \ref{tab:shapley_ML_full} ), the more important variables are: value added of industry, total retail sales of consumer goods, services PMI, steel production, and Korea's GDP growth rate.

\vspace*{\fill}\begin{center}
\begin{table}[ht]
    \scriptsize 
    \centering
    \resizebox{\textwidth}{!}{%
        \begin{tabularx}{\textwidth}{c|*{8}{>{\centering\arraybackslash}X}}
            \hline
            Variables & \makecell{AR} & \makecell{GBDT\\\_AE} & \makecell{GBDT\\\_HUBER} & \makecell{GBDT\\\_SB} & \makecell{RF\\\_AE} & \makecell{RF\\\_SE} & \makecell{XGB\\\_GBLINE\\AR} & \makecell{XGB\\\_GBTREE} \\
            \hline
            Industrial VA & \cellcolor[HTML]{BFBFBF}0.92 & \cellcolor[HTML]{BFBFBF}0.51 & \cellcolor[HTML]{BFBFBF}0.79 & \cellcolor[HTML]{BFBFBF}0.63 & \cellcolor[HTML]{BFBFBF}1.05 & \cellcolor[HTML]{BFBFBF}0.96 & \cellcolor[HTML]{BFBFBF}0.70 & \cellcolor[HTML]{BFBFBF}0.77 \\
            Imports & 0.33 & 0.03 & 0.01 & 0.01 & 0.07 & 0.01 & \cellcolor[HTML]{BFBFBF}0.58 & 0.10 \\
            Exports & 0.01 & 0.02 & 0.03 & 0.03 & 0.02 & 0.04 & 0.10 & 0.07 \\
            Retail Sales & \cellcolor[HTML]{BFBFBF}0.44 & \cellcolor[HTML]{BFBFBF}0.54 & \cellcolor[HTML]{BFBFBF}0.33 & \cellcolor[HTML]{BFBFBF}0.23 & \cellcolor[HTML]{BFBFBF}0.60 & \cellcolor[HTML]{BFBFBF}0.57 & \cellcolor[HTML]{BFBFBF}0.39 & \cellcolor[HTML]{BFBFBF}0.58 \\
            PMI Manu. & 0.23 & 0.17 & 0.01 & 0.03 & 0.06 & 0.07 & 0.02 & \cellcolor[HTML]{BFBFBF}0.20 \\
            PMI Manu. Prod. & \cellcolor[HTML]{BFBFBF}0.77 & 0.05 & 0.07 & 0.02 & 0.07 & \cellcolor[HTML]{BFBFBF}0.18 & 0.03 & 0.10 \\
            Sales-output Ratio & 0.02 & 0.03 & 0.02 & 0.03 & 0.03 & 0.06 & 0.05 & 0.06 \\
            Caixin PMI: Manu. & 0.07 & 0.07 & \cellcolor[HTML]{BFBFBF}0.13 & 0.06 & 0.03 & 0.06 & 0.01 & 0.07 \\
            USA GDP & \cellcolor[HTML]{BFBFBF}0.44 & 0.01 & 0.01 & 0.00 & 0.01 & 0.01 & 0.32 & 0.01 \\
            Japan GDP & 0.41 & 0.02 & 0.03 & 0.01 & 0.04 & 0.03 & 0.12 & 0.04 \\
            Eurozone GDP & 0.35 & 0.02 & 0.02 & 0.03 & 0.05 & 0.04 & 0.16 & 0.03 \\
            Korea GDP & 0.33 & \cellcolor[HTML]{BFBFBF}0.21 & \cellcolor[HTML]{BFBFBF}0.19 & 0.01 & \cellcolor[HTML]{BFBFBF}0.28 & 0.09 & 0.24 & 0.01 \\
            Exp. Delivery & 0.35 & 0.04 & 0.02 & 0.00 & 0.06 & 0.09 & \cellcolor[HTML]{BFBFBF}0.42 & 0.11 \\
            Freight Traffic & 0.13 & 0.05 & 0.08 & 0.08 & 0.18 & 0.13 & 0.29 & 0.10 \\
            Output: Steel & 0.13 & \cellcolor[HTML]{BFBFBF}0.33 & 0.13 & \cellcolor[HTML]{BFBFBF}0.29 & \cellcolor[HTML]{BFBFBF}0.38 & \cellcolor[HTML]{BFBFBF}0.45 & \cellcolor[HTML]{BFBFBF}0.37 & \cellcolor[HTML]{BFBFBF}0.61 \\
            Output: Electricity & 0.19 & 0.08 & 0.08 & 0.04 & 0.09 & 0.07 & 0.07 & 0.07 \\
            Floor Space: New & 0.26 & 0.07 & 0.05 & 0.03 & 0.07 & 0.03 & 0.21 & 0.04 \\
            Floor Space: Sold & 0.05 & 0.10 & 0.05 & \cellcolor[HTML]{BFBFBF}0.08 & 0.08 & 0.13 & 0.13 & 0.17 \\
            Fixed Asset Inv. & 0.05 & 0.03 & 0.08 & 0.04 & 0.04 & 0.05 & 0.10 & 0.14 \\
            Caixin PMI: Ser. & \cellcolor[HTML]{BFBFBF}0.54 & \cellcolor[HTML]{BFBFBF}0.34 & \cellcolor[HTML]{BFBFBF}0.23 & \cellcolor[HTML]{BFBFBF}0.16 & \cellcolor[HTML]{BFBFBF}0.42 & \cellcolor[HTML]{BFBFBF}0.25 & 0.15 & \cellcolor[HTML]{BFBFBF}0.24 \\
            \hline
        \end{tabularx}
    }
    \caption{Shapley Values for Machine Learning Models and AR Models (2006-2023)}
    \label{tab:shapley_ML_full} 
\end{table}
\end{center}\vspace*{\fill}

\vspace*{\fill}\begin{center}
\begin{table}[ht]
    \scriptsize
    \centering
    \resizebox{\textwidth}{!}{%
        \begin{tabularx}{\textwidth}{c|ccccccccccc}
            \hline
            Variables & \makecell{FM\\\_AR\\\_SB} & \makecell{FM\\\_GBDT\\\_AE} & \makecell{FM\\\_GBDT\\\_HUBER} & \makecell{FM\\\_GBDT\\\_SE} & \makecell{FM\\\_KRR\\\_POLY} & \makecell{FM\\\_KRR\\\_RBF} & \makecell{FM\\\_LASSO} & \makecell{FM\\\_RF\\\_AE} & \makecell{FM\\\_RF\\\_SB} & \makecell{FM\\\_XGB\\\_GBLINEAR} & \makecell{FM\\\_XGB\\\_GBTREE} \\
            \hline
            Industrial VA & 0.05 & 0.02 & 0.03 & 0.03 & 0.05 & 0.05 & 0.04 & 0.05 & 0.06 & 0.05 & 0.06 \\
            Imports & \cellcolor[HTML]{BFBFBF}0.47 & \cellcolor[HTML]{BFBFBF}0.15 & \cellcolor[HTML]{BFBFBF}0.18 & \cellcolor[HTML]{BFBFBF}0.17 & \cellcolor[HTML]{BFBFBF}0.48 & \cellcolor[HTML]{BFBFBF}0.48 & 0.34 & 0.36 & \cellcolor[HTML]{BFBFBF}0.47 & \cellcolor[HTML]{BFBFBF}0.47 & \cellcolor[HTML]{BFBFBF}0.39 \\
            Exports & 0.40 & 0.11 & 0.11 & 0.11 & 0.09 & 0.08 & 0.05 & 0.06 & 0.06 & 0.05 & 0.07 \\
            Retail Sales & 0.04 & 0.01 & 0.02 & 0.02 & 0.04 & 0.04 & 0.02 & 0.03 & 0.04 & 0.03 & 0.04 \\
            PMI Manu. & 0.00 & 0.00 & 0.01 & 0.01 & 0.01 & 0.01 & 0.00 & 0.00 & 0.01 & 0.01 & 0.01 \\
            PMI Manu. Prod. & 0.00 & 0.00 & 0.01 & 0.01 & 0.01 & 0.01 & 0.01 & 0.01 & 0.02 & 0.02 & 0.01 \\
            Sales-output Ratio & 0.01 & 0.01 & 0.01 & 0.01 & 0.00 & 0.00 & 0.00 & 0.00 & 0.00 & 0.00 & 0.00 \\
            Caixin PMI: Manu. & 0.00 & 0.01 & 0.01 & 0.01 & 0.00 & 0.00 & 0.00 & 0.01 & 0.01 & 0.01 & 0.01 \\
            USA GDP & 0.00 & 0.00 & 0.00 & 0.00 & 0.00 & 0.00 & 0.00 & 0.00 & 0.00 & 0.00 & 0.00 \\
            Japan GDP & 0.00 & 0.00 & 0.00 & 0.00 & 0.00 & 0.00 & 0.00 & 0.00 & 0.00 & 0.00 & 0.00 \\
            Eurozone GDP & 0.00 & 0.00 & 0.00 & 0.00 & 0.00 & 0.00 & 0.00 & 0.00 & 0.00 & 0.00 & 0.00 \\
            Korea GDP & 0.00 & 0.00 & 0.00 & 0.00 & 0.00 & 0.00 & 0.00 & 0.00 & 0.00 & 0.00 & 0.00 \\
            Exp. Delivery & 0.13 & 0.05 & 0.09 & 0.11 & 0.13 & 0.13 & 0.08 & 0.08 & 0.09 & 0.09 & 0.11 \\
            Freight Traffic & 0.05 & 0.01 & 0.02 & 0.02 & 0.02 & 0.02 & 0.01 & 0.01 & 0.01 & 0.01 & 0.02 \\
            Output: Steel & 0.00 & 0.01 & 0.02 & 0.02 & 0.02 & 0.02 & 0.01 & 0.01 & 0.01 & 0.01 & 0.02 \\
            Output: Electricity & 0.04 & 0.01 & 0.02 & 0.02 & 0.04 & 0.04 & 0.02 & 0.02 & 0.04 & 0.04 & 0.04 \\
            Floor Space: New & \cellcolor[HTML]{BFBFBF}0.76 & 0.13 & 0.14 & 0.14 & 0.36 & 0.36 & 0.34 & 0.36 & \cellcolor[HTML]{BFBFBF}0.64 & \cellcolor[HTML]{BFBFBF}0.61 & \cellcolor[HTML]{BFBFBF}0.64 \\
            Floor Space: Sold & 0.30 & 0.18 & 0.21 & 0.20 & 0.30 & 0.30 & 0.29 & 0.29 & \cellcolor[HTML]{BFBFBF}0.38 & \cellcolor[HTML]{BFBFBF}0.37 & \cellcolor[HTML]{BFBFBF}0.38 \\
            Fixed Asset Inv. & \cellcolor[HTML]{BFBFBF}0.19 & \cellcolor[HTML]{BFBFBF}0.13 & \cellcolor[HTML]{BFBFBF}0.16 & \cellcolor[HTML]{BFBFBF}0.16 & 0.22 & 0.22 & 0.18 & 0.19 & 0.34 & \cellcolor[HTML]{BFBFBF}0.32 & \cellcolor[HTML]{BFBFBF}0.34 \\
            Caixin PMI: Ser. & \cellcolor[HTML]{BFBFBF}0.05 & 0.02 & 0.04 & 0.04 & \cellcolor[HTML]{BFBFBF}0.05 & \cellcolor[HTML]{BFBFBF}0.05 & 0.03 & 0.04 & 0.06 & \cellcolor[HTML]{BFBFBF}0.05 & \cellcolor[HTML]{BFBFBF}0.05 \\
            \hline
        \end{tabularx}
    }
    \caption{Shapley Values for Factor Models and FM-ML Coupled Models (2006-2023)}
    \label{tab:sharpleyFMfull} 
\end{table}
\end{center}\vspace*{\fill}

Tables \ref{tab:sharpleyML_2008-2010} through \ref{tab:sharpleyFM2020-2022} show the Shapley values for each model during the global financial crisis (2008-2010) and the COVID-19 pandemic (2020-2022). Overall, while there may be slight differences in the ranking of Shapley values between models, the performance remains relatively consistent over time. An exception is observed during the COVID-19 pandemic, where changes in freight volume had a stronger impact on the machine learning models' results. During this period, the correlation between freight volume changes and economic growth increased due to the pandemic's effect on goods transportation. Outside of this period, freight volume does not appear as a more important driver in terms of Shapley value for explaining GDP changes.

\vspace*{\fill}\begin{center}
\begin{table}[ht]
    \scriptsize
    \centering
    \resizebox{\textwidth}{!}{%
        \begin{tabularx}{\textwidth}{c|*{8}{>{\centering\arraybackslash}X}}
            \hline
            Variables & \makecell{AR} & \makecell{GBDT\\\_AE} & \makecell{GBDT\\\_HUBER} & \makecell{GBDT\\\_SB} & \makecell{RF\\\_AE} & \makecell{RF\\\_SE} & \makecell{XGB\\\_GBLINE\\AR} & \makecell{XGB\\\_GBTREE} \\
            \hline
            Industrial VA & 1.01 & 0.60 & 0.79 & 0.58 & \cellcolor[HTML]{BFBFBF}1.27 & \cellcolor[HTML]{BFBFBF}0.99 & \cellcolor[HTML]{BFBFBF}0.76 & \cellcolor[HTML]{BFBFBF}0.85 \\
            Imports & \cellcolor[HTML]{BFBFBF}0.59 & 0.02 & 0.01 & 0.02 & 0.09 & 0.01 & 0.34 & 0.13 \\
            Exports & 0.01 & 0.02 & 0.04 & 0.05 & 0.01 & 0.05 & 0.15 & 0.08 \\
            Retail Sales & \cellcolor[HTML]{BFBFBF}0.63 & \cellcolor[HTML]{BFBFBF}0.51 & \cellcolor[HTML]{BFBFBF}0.29 & \cellcolor[HTML]{BFBFBF}0.24 & \cellcolor[HTML]{BFBFBF}0.65 & \cellcolor[HTML]{BFBFBF}0.60 & 0.42 & \cellcolor[HTML]{BFBFBF}0.73 \\
            PMI Manu. & \cellcolor[HTML]{BFBFBF}0.45 & 0.21 & 0.02 & 0.03 & 0.10 & 0.08 & 0.05 & 0.26 \\
            PMI Manu. Prod. & \cellcolor[HTML]{BFBFBF}1.46 & 0.06 & 0.09 & 0.02 & 0.12 & 0.22 & 0.10 & 0.20 \\
            Sales-output Ratio & 0.02 & 0.02 & 0.01 & 0.01 & 0.01 & 0.03 & 0.05 & 0.03 \\
            Caixin PMI: Manu. & 0.15 & 0.05 & \cellcolor[HTML]{BFBFBF}0.25 & 0.01 & 0.06 & 0.06 & 0.01 & 0.09 \\
            USA GDP & \cellcolor[HTML]{BFBFBF}0.80 & 0.01 & 0.00 & 0.00 & 0.01 & 0.00 & 0.29 & 0.01 \\
            Japan GDP & \cellcolor[HTML]{BFBFBF}0.86 & 0.04 & 0.04 & 0.01 & 0.05 & 0.03 & 0.07 & 0.03 \\
            Eurozone GDP & 0.44 & 0.02 & 0.00 & 0.01 & 0.05 & 0.03 & 0.07 & 0.03 \\
            Korea GDP & \cellcolor[HTML]{BFBFBF}0.70 & \cellcolor[HTML]{BFBFBF}0.32 & \cellcolor[HTML]{BFBFBF}0.21 & \cellcolor[HTML]{BFBFBF}0.00 & \cellcolor[HTML]{BFBFBF}0.28 & \cellcolor[HTML]{BFBFBF}0.13 & 0.50 & 0.01 \\
            Exp. Delivery & \cellcolor[HTML]{BFBFBF}0.61 & 0.03 & 0.04 & 0.07 & 0.11 & 0.07 & \cellcolor[HTML]{BFBFBF}0.72 & 0.10 \\
            Freight Traffic & 0.12 & 0.03 & 0.03 & 0.02 & 0.10 & 0.05 & 0.20 & 0.12 \\
            Output:Steel & 0.18 & \cellcolor[HTML]{BFBFBF}0.35 & 0.18 & \cellcolor[HTML]{BFBFBF}0.33 & \cellcolor[HTML]{BFBFBF}0.43 & \cellcolor[HTML]{BFBFBF}0.54 & \cellcolor[HTML]{BFBFBF}0.70 & \cellcolor[HTML]{BFBFBF}0.73 \\
            Output: Electricity & \cellcolor[HTML]{BFBFBF}0.30 & 0.09 & 0.09 & 0.02 & 0.09 & 0.07 & 0.07 & 0.07 \\
            Floor Space: New & \cellcolor[HTML]{BFBFBF}0.42 & 0.09 & 0.06 & 0.03 & 0.09 & 0.05 & 0.34 & 0.04 \\
            Floor Space: Sold & 0.06 & 0.10 & 0.05 & \cellcolor[HTML]{BFBFBF}0.06 & 0.10 & 0.14 & 0.15 & 0.16 \\
            Fixed Asset Inv. & 0.05 & 0.03 & 0.06 & 0.02 & 0.04 & 0.05 & 0.08 & 0.10 \\
            Caixin PMI: Ser. & \cellcolor[HTML]{BFBFBF}0.60 & \cellcolor[HTML]{BFBFBF}0.34 & \cellcolor[HTML]{BFBFBF}0.20 & \cellcolor[HTML]{BFBFBF}0.17 & \cellcolor[HTML]{BFBFBF}0.39 & \cellcolor[HTML]{BFBFBF}0.18 & 0.16 & \cellcolor[HTML]{BFBFBF}0.25 \\
            \hline
        \end{tabularx}
    }
    \caption{Shapley Values for Machine Learning Models and AR Models (2008-2010)}
    \label{tab:sharpleyML_2008-2010} 
\end{table}
\vspace*{\fill}\end{center}

\vspace*{\fill}\begin{center}
\begin{table}[ht]
    \scriptsize
    \centering
    \resizebox{\textwidth}{!}{%
        \begin{tabularx}{\textwidth}{c|ccccccccccc}
            \hline
            Variables & \makecell{FM\\\_AR\\\_SB} & \makecell{FM\\\_GBDT\\\_AE} & \makecell{FM\\\_GBDT\\\_HUBER} & \makecell{FM\\\_GBDT\\\_SE} & \makecell{FM\\\_KRR\\\_POLY} & \makecell{FM\\\_KRR\\\_RBF} & \makecell{FM\\\_LASSO} & \makecell{FM\\\_RF\\\_AE} & \makecell{FM\\\_RF\\\_SB} & \makecell{FM\\\_XGB\\\_GBLINEAR} & \makecell{FM\\\_XGB\\\_GBTREE} \\
            \hline
            Industrial VA & 0.05 & 0.02 & \cellcolor[HTML]{BFBFBF}0.02 & 0.06 & 0.06 & 0.04 & 0.04 & 0.06 & 0.05 & 0.05 & 0.05 \\
            Imports & \cellcolor[HTML]{BFBFBF}0.84 & \cellcolor[HTML]{BFBFBF}0.24 & 0.26 & \cellcolor[HTML]{BFBFBF}0.27 & \cellcolor[HTML]{BFBFBF}0.85 & \cellcolor[HTML]{BFBFBF}0.61 & \cellcolor[HTML]{BFBFBF}0.74 & \cellcolor[HTML]{BFBFBF}0.70 & \cellcolor[HTML]{BFBFBF}0.84 & \cellcolor[HTML]{BFBFBF}0.60 \\
            Exports & \cellcolor[HTML]{BFBFBF}0.62 & \cellcolor[HTML]{BFBFBF}0.19 & \cellcolor[HTML]{BFBFBF}0.21 & 0.21 & 0.31 & 0.31 & 0.42 & 0.45 & \cellcolor[HTML]{BFBFBF}0.55 & \cellcolor[HTML]{BFBFBF}0.61 & \cellcolor[HTML]{BFBFBF}0.48 \\
            Retail Sales & 0.05 & 0.02 & 0.03 & 0.03 & 0.04 & 0.01 & 0.02 & 0.05 & 0.02 & 0.06 & 0.06 \\
            PMI Manu. & 0.09 & 0.00 & 0.01 & 0.01 & 0.01 & 0.01 & 0.01 & 0.01 & 0.01 & 0.00 & 0.01 \\
            PMI Manu. Prod. & 0.02 & 0.01 & 0.01 & 0.02 & 0.02 & 0.01 & 0.01 & 0.01 & 0.02 & 0.02 & 0.01 \\
            Sales-output Ratio & 0.01 & 0.01 & 0.01 & 0.01 & 0.01 & 0.01 & 0.01 & 0.00 & 0.01 & 0.00 & 0.01 \\
            Caixin PMI: Manu. & 0.00 & 0.00 & 0.00 & 0.01 & 0.01 & 0.01 & 0.01 & 0.01 & 0.01 & 0.01 & 0.01 \\
            USA GDP & 0.01 & 0.00 & 0.00 & 0.00 & 0.01 & 0.00 & 0.01 & 0.00 & 0.01 & 0.01 & 0.00 \\
            Japan GDP & 0.01 & 0.00 & 0.00 & 0.00 & 0.00 & 0.01 & 0.01 & 0.00 & 0.00 & 0.01 & 0.00 \\
            Eurozone GDP & 0.00 & 0.00 & 0.00 & 0.00 & 0.01 & 0.00 & 0.01 & 0.00 & 0.00 & 0.00 & 0.00 \\
            Korea GDP & 0.01 & 0.00 & 0.00 & 0.00 & 0.01 & 0.00 & 0.01 & 0.01 & 0.00 & 0.00 & 0.00 \\
            Exp. Delivery & 0.22 & 0.06 & 0.07 & 0.07 & 0.10 & 0.07 & \cellcolor[HTML]{BFBFBF}0.23 & 0.16 & 0.14 & \cellcolor[HTML]{BFBFBF}0.22 & 0.16 \\
            Freight Traffic & 0.05 & 0.01 & 0.03 & 0.02 & 0.03 & 0.02 & 0.06 & 0.07 & 0.07 & 0.04 & 0.06 \\
            Output: Steel & 0.10 & 0.02 & 0.03 & 0.03 & 0.05 & 0.03 & 0.05 & 0.06 & 0.07 & 0.05 & 0.07 \\
            Output: Electricity & \cellcolor[HTML]{BFBFBF}0.07 & 0.01 & 0.02 & 0.02 & 0.01 & 0.02 & 0.01 & 0.02 & 0.03 & 0.01 & \cellcolor[HTML]{BFBFBF}0.06 \\
            Floor Space: New & \cellcolor[HTML]{BFBFBF}0.40 & \cellcolor[HTML]{BFBFBF}0.12 & 0.11 & 0.09 & 0.11 & 0.09 & 0.12 & 0.14 & 0.15 & 0.17 & \cellcolor[HTML]{BFBFBF}0.36 \\
            Floor Space: Sold & 0.40 & 0.16 & 0.15 & 0.17 & 0.18 & 0.17 & \cellcolor[HTML]{BFBFBF}0.19 & 0.17 & \cellcolor[HTML]{BFBFBF}0.40 & 0.19 & 0.14 \\
            Fixed Asset Inv. & \cellcolor[HTML]{BFBFBF}0.19 & \cellcolor[HTML]{BFBFBF}0.16 & 0.16 & 0.17 & 0.17 & 0.19 & 0.18 & 0.18 & 0.17 & 0.16 & 0.15 \\
            Caixin PMI: Ser. & 0.06 & 0.02 & 0.03 & 0.03 & 0.04 & 0.02 & 0.07 & 0.06 & 0.06 & 0.04 & 0.07 \\
            \hline
        \end{tabularx}
    }
    \caption{Shapley Values for Factor Models and FM-ML Coupled Models (2008-2010)}
    \label{tab:sharpleyFM2008-2010} 
\end{table}
\vspace*{\fill}\end{center}

\vspace*{\fill}\begin{center}
\begin{table}[ht]
    \scriptsize 
    \centering
    \resizebox{\textwidth}{!}{%
        \begin{tabularx}{\textwidth}{c|*{8}{>{\centering\arraybackslash}X}}
            \hline
            Variables & \makecell{AR} & \makecell{GBDT\\\_AE} & \makecell{GBDT\\\_HUBER} & \makecell{GBDT\\\_SB} & \makecell{RF\\\_AE} & \makecell{RF\\\_SE} & \makecell{XGB\\\_GBLINE\\AR} & \makecell{XGB\\\_GBTREE} \\
            \hline
            Industrial VA & \cellcolor[HTML]{BFBFBF}1.38 & 0.55 & \cellcolor[HTML]{BFBFBF}1.22 & \cellcolor[HTML]{BFBFBF}1.07 & 0.97 & \cellcolor[HTML]{BFBFBF}1.25 & 1.05 & 1.01 \\
            Imports & 0.30 & 0.02 & 0.02 & 0.02 & 0.06 & 0.02 & \cellcolor[HTML]{BFBFBF}0.53 & 0.50 \\
            Exports & 0.01 & 0.05 & 0.02 & 0.02 & 0.06 & 0.04 & 0.11 & 0.09 \\
            Retail Sales & 1.01 & 0.26 & 0.85 & 0.93 & \cellcolor[HTML]{BFBFBF}1.43 & \cellcolor[HTML]{BFBFBF}1.30 & \cellcolor[HTML]{BFBFBF}1.30 & 0.18 \\
            PMI Manu. & 0.21 & 0.01 & 0.01 & 0.01 & 0.01 & 0.06 & 0.05 & 0.01 \\
            PMI Manu. Prod. & 0.86 & 0.05 & 0.09 & 0.07 & 0.07 & \cellcolor[HTML]{BFBFBF}0.19 & 0.21 & 0.18 \\
            Sales-output Ratio & 0.03 & 0.04 & 0.03 & 0.04 & 0.05 & 0.06 & 0.08 & 0.05 \\
            Caixin PMI: Manu. & 0.06 & 0.03 & 0.03 & 0.01 & 0.03 & 0.05 & 0.05 & 0.03 \\
            USA GDP & 0.05 & 0.00 & 0.00 & 0.00 & 0.01 & 0.01 & 0.10 & 0.03 \\
            Japan GDP & \cellcolor[HTML]{BFBFBF}0.62 & 0.01 & 0.01 & 0.01 & 0.05 & 0.03 & 0.02 & 0.01 \\
            Eurozone GDP & \cellcolor[HTML]{BFBFBF}0.86 & 0.02 & 0.01 & 0.01 & 0.07 & 0.02 & 0.01 & 0.03 \\
            Korea GDP & 0.92 & 0.04 & 0.02 & 0.01 & 0.09 & 0.05 & 0.01 & 0.03 \\
            Exp. Delivery & 0.26 & 0.04 & 0.04 & 0.05 & 0.06 & 0.06 & \cellcolor[HTML]{BFBFBF}0.31 & 0.30 \\
            Freight Traffic & 0.29 & 0.02 & 0.01 & 0.01 & 0.04 & 0.01 & \cellcolor[HTML]{BFBFBF}0.64 & 0.21 \\
            Output: Steel & 0.14 & \cellcolor[HTML]{BFBFBF}0.40 & 0.08 & 0.08 & \cellcolor[HTML]{BFBFBF}0.51 & 0.05 & \cellcolor[HTML]{BFBFBF}0.41 & \cellcolor[HTML]{BFBFBF}0.91 \\
            Output: Electricity & 0.17 & 0.12 & 0.10 & 0.09 & 0.03 & 0.09 & 0.02 & 0.03 \\
            Floor Space: New & 0.13 & 0.05 & 0.06 & 0.08 & 0.07 & 0.06 & \cellcolor[HTML]{BFBFBF}1.24 & 0.36 \\
            Floor Space: Sold & 0.06 & 0.17 & 0.08 & 0.11 & 0.15 & 0.10 & \cellcolor[HTML]{BFBFBF}0.23 & 0.13 \\
            Fixed Asset Inv. & 0.19 & 0.16 & 0.08 & 0.12 & 0.18 & 0.20 & \cellcolor[HTML]{BFBFBF}0.24 & \cellcolor[HTML]{BFBFBF}0.25 \\
            Caixin PMI: Ser. & \cellcolor[HTML]{BFBFBF}0.50 & \cellcolor[HTML]{BFBFBF}0.30 & \cellcolor[HTML]{BFBFBF}0.23 & \cellcolor[HTML]{BFBFBF}0.12 & \cellcolor[HTML]{BFBFBF}0.73 & \cellcolor[HTML]{BFBFBF}0.29 & 0.14 & \cellcolor[HTML]{BFBFBF}0.28 \\
            \hline
        \end{tabularx}
    }
    \caption{Shapley Values for Machine Learning Models and AR Models (2020-2022)}
    \label{tab:sharpleyML_2020-2022} 
\end{table}
\vspace*{\fill}\end{center}

\vspace*{\fill}\begin{center}
\begin{table}[ht]
    \scriptsize
    \centering
    \resizebox{\textwidth}{!}{%
        \begin{tabularx}{\textwidth}{c|ccccccccccc}
            \hline
            Variables & \makecell{FM\\\_AR\\\_SB} & \makecell{FM\\\_GBDT\\\_AE} & \makecell{FM\\\_GBDT\\\_HUBER} & \makecell{FM\\\_GBDT\\\_SE} & \makecell{FM\\\_KRR\\\_POLY} & \makecell{FM\\\_KRR\\\_RBF} & \makecell{FM\\\_LASSO} & \makecell{FM\\\_RF\\\_AE} & \makecell{FM\\\_RF\\\_SB} & \makecell{FM\\\_XGB\\\_GBLINEAR} & \makecell{FM\\\_XGB\\\_GBTREE} \\
            \hline
            Industrial VA & 0.07 & 0.02 & 0.05 & 0.04 & 0.08 & 0.08 & 0.06 & 0.07 & 0.11 & 0.07 & 0.11 \\
            Imports & \cellcolor[HTML]{BFBFBF}0.43 & \cellcolor[HTML]{BFBFBF}0.11 & \cellcolor[HTML]{BFBFBF}0.17 & \cellcolor[HTML]{BFBFBF}0.16 & \cellcolor[HTML]{BFBFBF}0.45 & \cellcolor[HTML]{BFBFBF}0.44 & \cellcolor[HTML]{BFBFBF}0.31 & \cellcolor[HTML]{BFBFBF}0.28 & \cellcolor[HTML]{BFBFBF}0.44 & \cellcolor[HTML]{BFBFBF}0.43 & \cellcolor[HTML]{BFBFBF}0.33 \\
            Exports & 0.45 & 0.10 & 0.17 & 0.17 & 0.52 & 0.51 & 0.31 & 0.32 & 0.32 & 0.35 & 0.35 \\
            Retail Sales & 0.08 & 0.02 & 0.05 & 0.04 & 0.09 & 0.09 & 0.07 & 0.06 & 0.10 & 0.08 & 0.10 \\
            PMI Manu. & 0.00 & 0.01 & 0.00 & 0.01 & 0.00 & 0.00 & 0.01 & 0.01 & 0.01 & 0.00 & 0.02 \\
            PMI Manu. Prod. & 0.01 & 0.00 & 0.01 & 0.01 & 0.01 & 0.01 & 0.01 & 0.01 & 0.01 & 0.01 & 0.02 \\
            Sales-output Ratio & 0.00 & 0.00 & 0.00 & 0.00 & 0.00 & 0.00 & 0.00 & 0.00 & 0.00 & 0.00 & 0.00 \\
            Caixin PMI: Manu. & 0.00 & 0.00 & 0.00 & 0.00 & 0.00 & 0.00 & 0.00 & 0.00 & 0.00 & 0.00 & 0.01 \\
            USA GDP & 0.00 & 0.00 & 0.00 & 0.00 & 0.00 & 0.00 & 0.00 & 0.00 & 0.00 & 0.00 & 0.00 \\
            Japan GDP & 0.00 & 0.00 & 0.00 & 0.00 & 0.00 & 0.00 & 0.00 & 0.00 & 0.00 & 0.00 & 0.00 \\
            Eurozone GDP & 0.00 & 0.00 & 0.00 & 0.00 & 0.00 & 0.00 & 0.00 & 0.00 & 0.00 & 0.00 & 0.00 \\
            Korea GDP & 0.01 & 0.00 & 0.00 & 0.00 & 0.01 & 0.01 & 0.00 & 0.01 & 0.01 & 0.01 & 0.00 \\
            Exp. Delivery & 0.09 & 0.02 & 0.07 & 0.04 & 0.04 & 0.05 & 0.06 & 0.09 & 0.13 & 0.09 & 0.10 \\
            Freight Traffic & 0.11 & 0.02 & 0.02 & 0.05 & 0.05 & 0.04 & 0.07 & 0.09 & 0.09 & 0.07 & 0.08 \\
            Output: Steel & 0.08 & 0.01 & 0.03 & 0.05 & 0.07 & 0.05 & 0.05 & 0.08 & 0.11 & 0.08 & 0.10 \\
            Output: Electricity & \cellcolor[HTML]{BFBFBF}0.17 & 0.13 & 0.10 & 0.09 & 0.12 & 0.07 & 0.09 & 0.07 & 0.04 & 0.04 & 0.04 \\
            Floor Space: New & \cellcolor[HTML]{BFBFBF}0.98 & \cellcolor[HTML]{BFBFBF}0.13 & 0.54 & 0.40 & 0.66 & 0.65 & 0.97 & 0.80 & \cellcolor[HTML]{BFBFBF}0.98 & 1.03 \\
            Floor Space: Sold & \cellcolor[HTML]{BFBFBF}0.39 & 0.16 & 0.20 & 0.18 & \cellcolor[HTML]{BFBFBF}0.49 & 0.38 & 0.38 & 0.31 & 0.38 & 0.37 \\
            Fixed Asset Inv. & \cellcolor[HTML]{BFBFBF}0.45 & 0.28 & 0.15 & 0.15 & \cellcolor[HTML]{BFBFBF}0.62 & \cellcolor[HTML]{BFBFBF}0.57 & 0.13 & 0.31 & \cellcolor[HTML]{BFBFBF}0.44 & 0.45 \\
            Caixin PMI: Ser. & 0.05 & 0.01 & 0.04 & 0.06 & 0.06 & 0.06 & 0.03 & 0.09 & 0.05 & 0.08 \\
            \hline
        \end{tabularx}
    }
    \caption{Shapley Values for Factor Models and FM-ML Coupled Models (2020-2022)}
    \label{tab:sharpleyFM2020-2022} 
\end{table}
\vspace*{\fill}\end{center}

\subsection{Local Perspective}

Global analysis conveys the importance of a variable within the test set, while local imputation assesses a variable's significance for predicted values within a specific period.

Figures X1 to X4 illustrate the Shapley value function forms for three machine learning models and three factor-and-machine learning coupled models. The curves represent the fitted results. We selected the five most important variables for each set of models based on global results, totaling ten variables: value added of industry, total retail sales of consumer goods, PMI of the service industry, steel production, GDP growth rate of South Korea, area of new housing construction, area of sales of commercial properties, amount of imports, amount of exports, and amount of completed investment in non-private fixed assets. Rows represent variables, and columns represent different models. Each graph depicts local Shapley values (vertical axis) against observed variable input values (horizontal axis). Although the four machine learning methods employ different learning mechanisms and differ in global feature importance, the functional forms derived from the models' variable learning are consistent, indicating the robustness of the learned functions.

\begin{figure}
\centering
\includegraphics[width=0.7\textwidth]{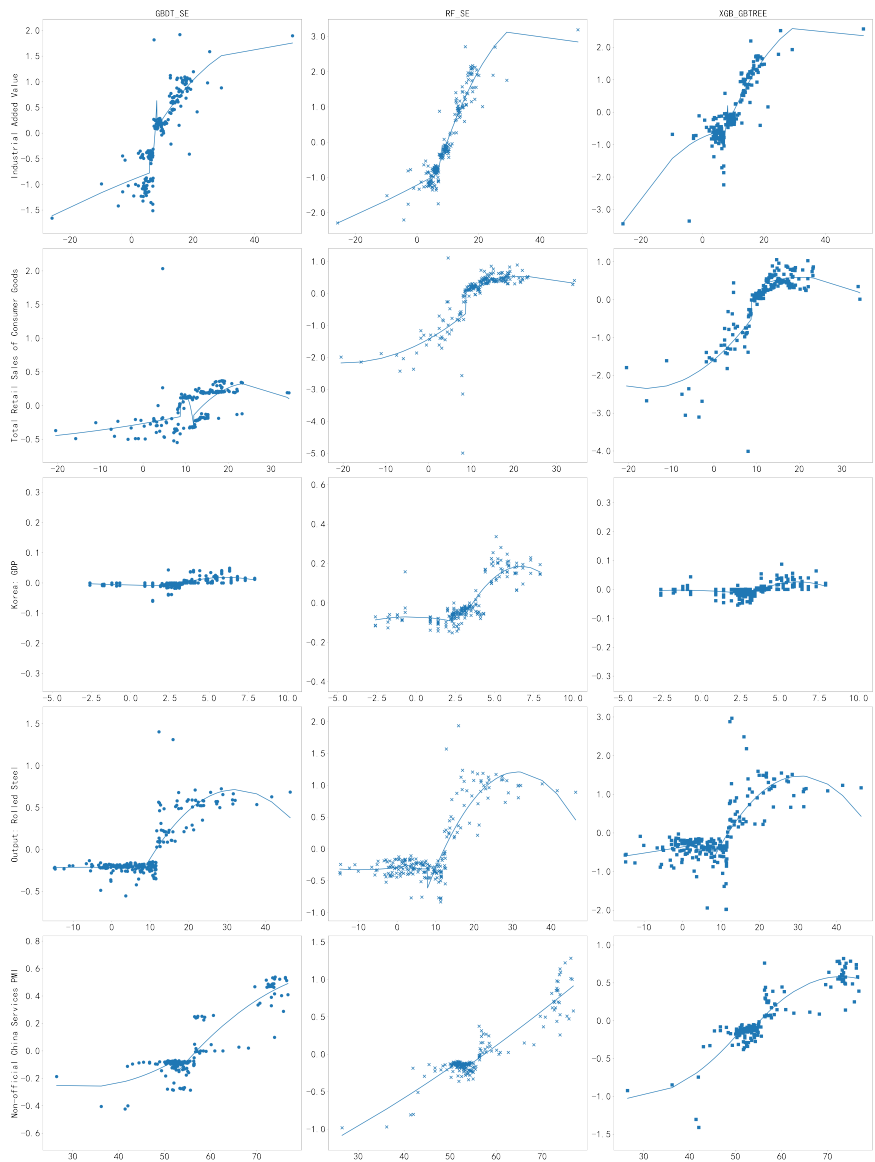}
\caption{ML Model Comparisons}
\label{fig:all_nonFM_part1}
\end{figure}

\begin{figure}
\centering
\includegraphics[width=0.7\textwidth]{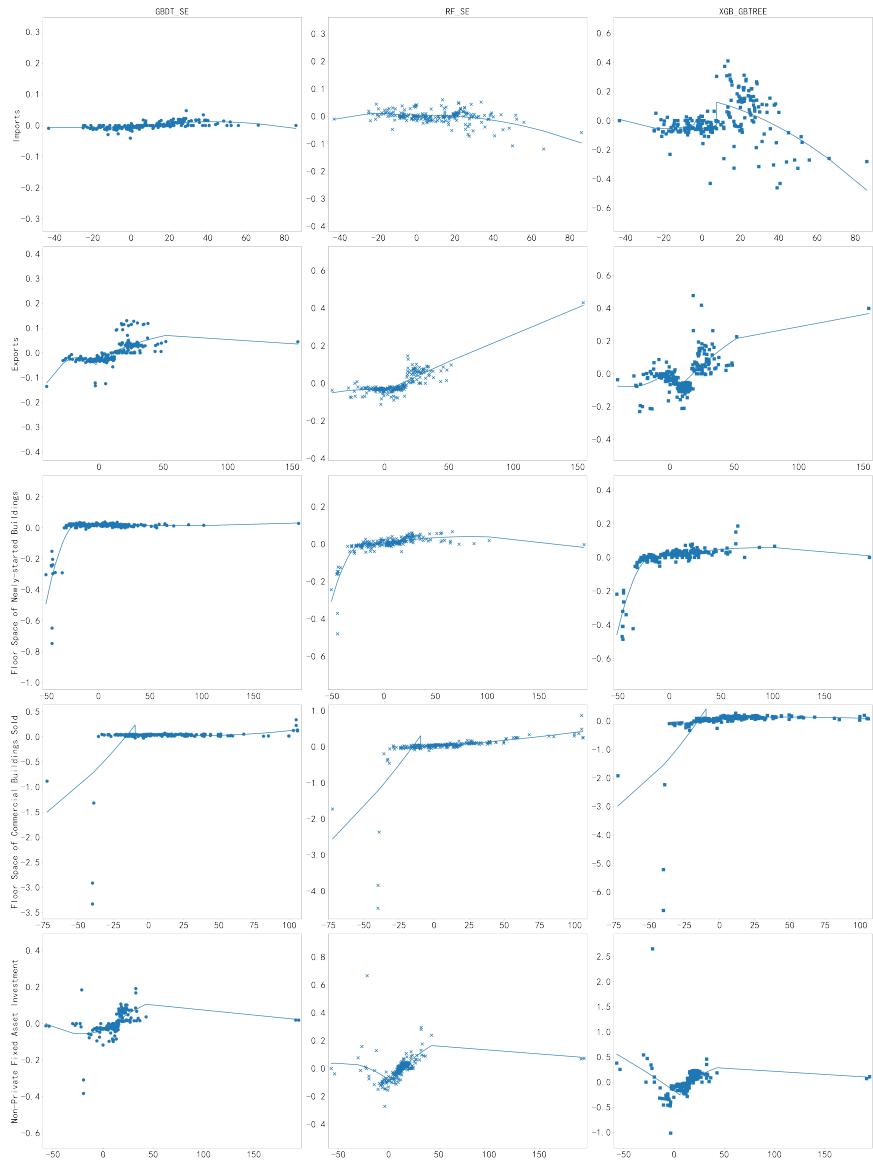}
\caption{ML Model Comparisons (continued)}
\label{fig:all_nonFM_part2}
\end{figure}

\begin{figure}
\centering
\includegraphics[width=0.7\textwidth]{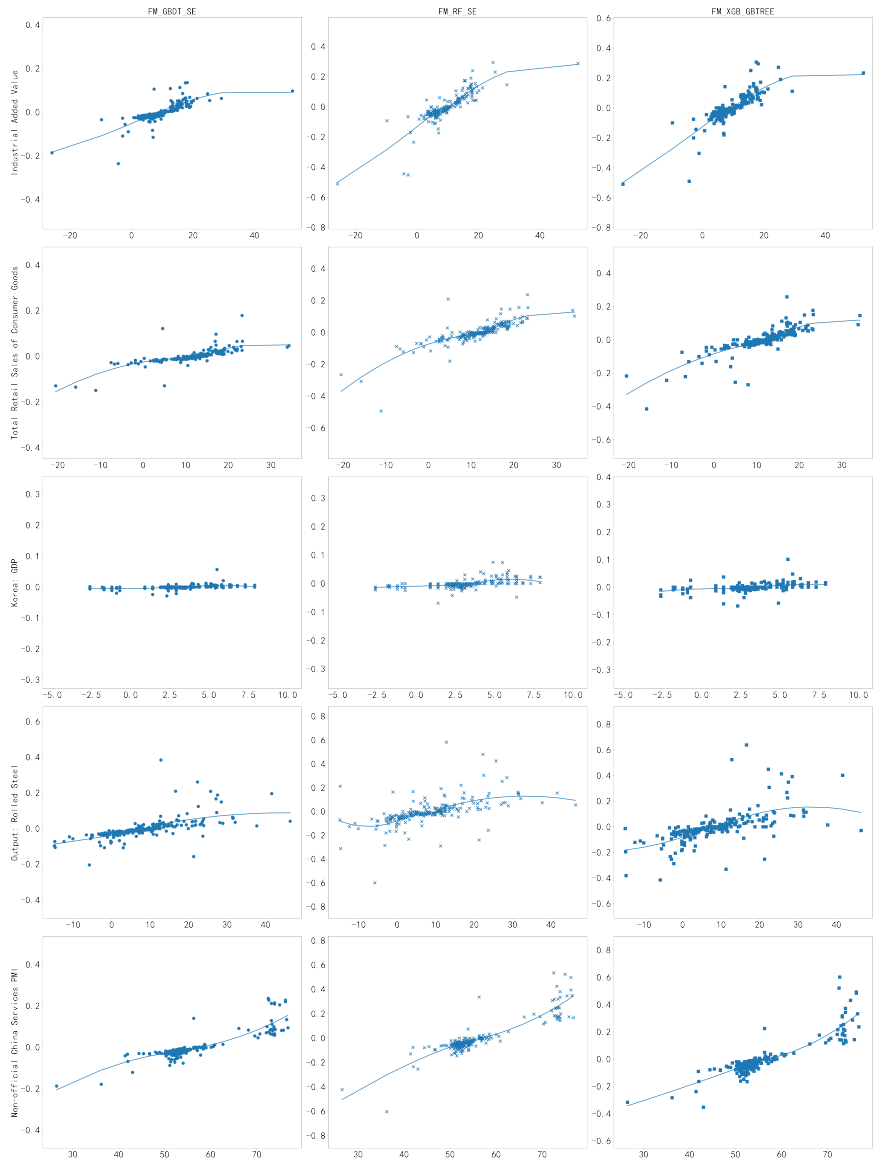}
\caption{FM Models and Combined Models}
\label{fig:all_FM_part1}
\end{figure}

\begin{figure}
\centering
\includegraphics[width=0.7\textwidth]{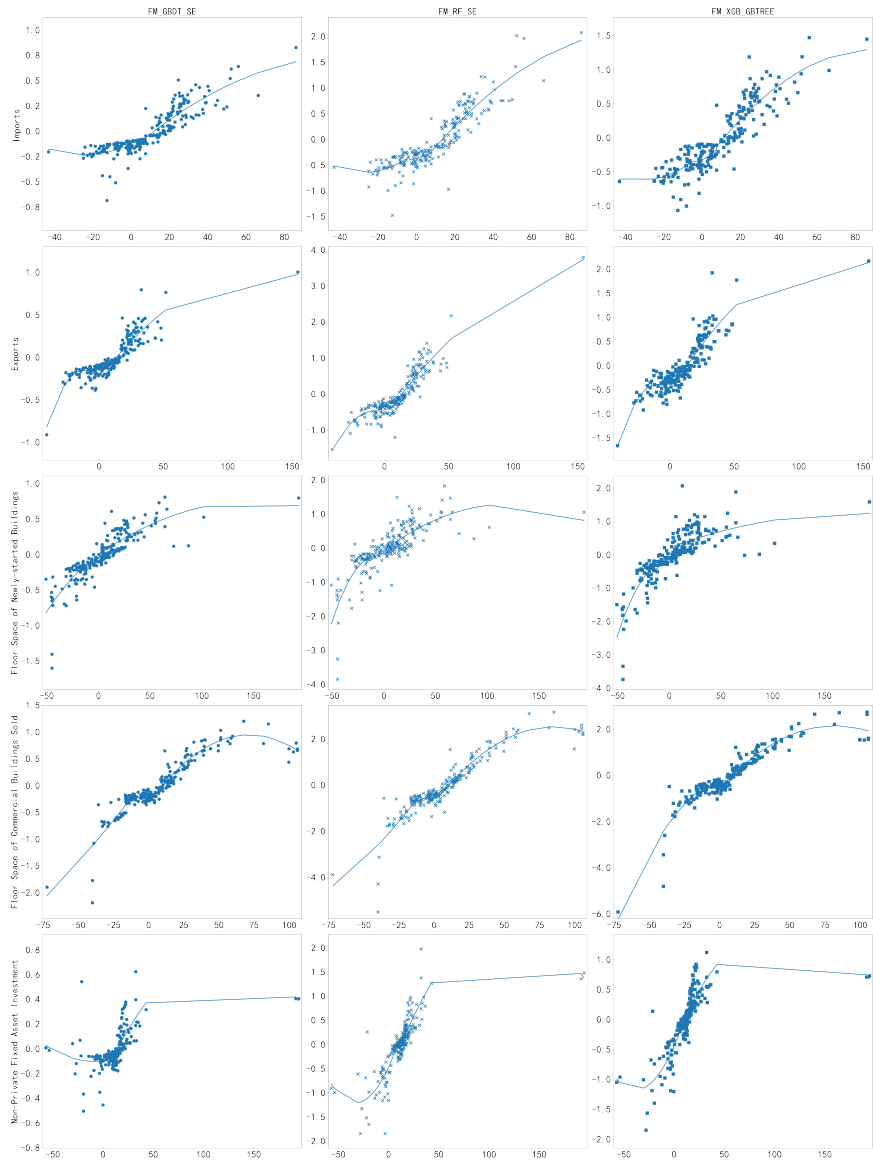}
\caption{FM Models and Combined Models (continued)}
\label{fig:all_FM_part2}
\end{figure}

\begin{figure}
\centering
\includegraphics[width=0.7\textwidth]{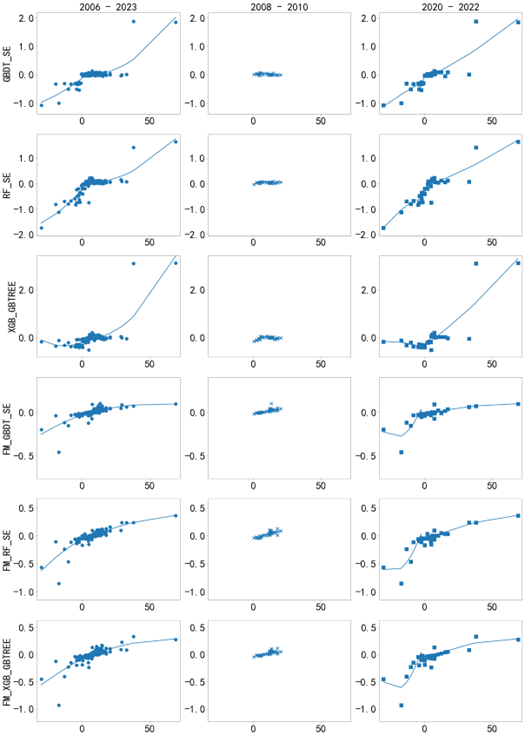}
\caption{Comparison of the functional form of freight volumes over different periods of time for each model}
\label{fig:Freight}
\end{figure}

Figure \ref{fig:Freight} compares the differences in the functional form fitted by the models to the variable change in freight volume over time. During the 2020-2022 COVID-19 pandemic period, the average Shapley values of the three machine learning models were higher, reflecting the impact of the epidemic on production and consumption. In contrast, during the same period, the combined model of factor and machine learning models did not show higher Shapley values. Additionally, during the global financial crisis from 2008 to 2010, the change in freight volume did not have a significant impact on GDP forecasting results, with the Shapley values of all models remaining near zero.

\section{Robustness Check}

\subsection{Evaluation Method}

The empirical results in this paper compare the accuracy of various machine learning and combined model forecasts using root mean square error (RMSE) and mean absolute error (MAE). To investigate whether out-of-sample prediction results depend on the assessment method, we use the AR model in the econometric model as the benchmark prediction. We then compare whether the machine learning model and combined model significantly improve their predictions relative to the benchmark model, thereby testing the robustness of the out-of-sample prediction results.

\begin{figure}
\centering
\includegraphics[width=0.8\textwidth]{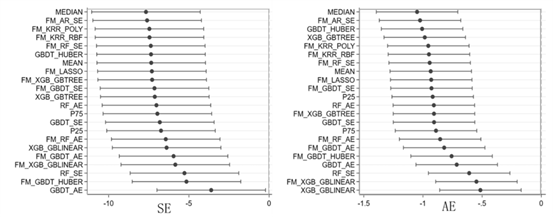}
\caption{Comparison of Machine Learning Models to Benchmark Models, 2010-2023}
\label{fig:SE_AE1}
\end{figure}

\begin{figure}
\centering
\includegraphics[width=0.8\textwidth]{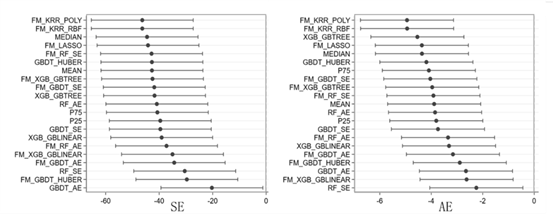}
\caption{Comparison of Machine Learning Models and Benchmark Models during Covid-19 Period, 2020-2022}
\label{fig:SE_AE2}
\end{figure}

Figures\ref{fig:SE_AE1} and \ref{fig:SE_AE2} show the out-of-sample performance of the other models relative to the AR model for the 2010-2023 and 2020-2022 COVID-19 pandemic periods, respectively. The values on the horizontal axis in both figures represent the model fixed effects in the following panel regressions:

\begin{equation}
    {\mathrm{\ eval\_metric}}_{t,h,m}=\alpha_m+\gamma_{t,h}+{\grave{o}}_{t,h,m},
\end{equation}
where \(\alpha_m\) represents the model fixed effect and \(\gamma_{t,h}\) denotes the fixed effect of time and prediction step. We use squared error (SE) and absolute error (AE) as the measures of \(\mathrm{eval\_metric}_{t,h,m}\).

A comparison of the two figures shows that the prediction performance of all types of machine learning models significantly improves compared to the benchmark model. However, the SE and AE during the COVID-19 pandemic period are significantly higher than during periods of stable economic operation. Throughout the 2010-2022 period, the median prediction accuracy of all machine learning-related models is higher, while kernel ridge regression associated with machine learning models demonstrates higher prediction accuracy during the COVID-19 pandemic. These conclusions align with the main findings in the empirical results section, indicating that the out-of-sample prediction results in this paper do not depend on the error analysis method chosen and that the out-of-sample prediction performance is robust.

\subsection{Testing Window}
To test the robustness of the empirical results against the selection of the evaluation window time, we select additional historical periods to evaluate model performance. Table \ref{tab:ML1996-2009} presents the forecasting performance of the econometric model, the machine learning model, the combined model, the mixed model and the weighted model in predicting China's quarterly GDP growth rate for the periods 1996-1999, 2000-2003, and 2004-2009. It also highlights the model's performance during the 1997-1998 Asian financial crisis. In this context, mean G2 and median G2 represent the mean and median of the machine learning models, respectively; mean G3 and median G3 represent the mean and median of the combined models, respectively; and mean G2\&3 and median G2\&3 represent the mean and median of both machine learning and combined models.

Table \ref{tab:ML1996-2009} shows that the machine learning models outperform the benchmark econometric models in terms of forecasting performance in the 1996-2009 period. During the 1997-1998 Asian financial crisis period, the difference between the extremes in the training data and the extremes in the prediction intervals is small, so the ridge regression (KRR) correlation model does not outperform the other machine learning models, which is in line with the results we obtained in Table 7. Overall, the predictive performance of the models in Table 9 is consistent with the model performance for the period 2005-2022 presented in Tables 6 and 7, indicating that the out-of-sample prediction results in this paper do not depend on the selection of the evaluation window time. The out-of-sample forecast performance is robust.

Table \ref{tab:ML1996-2009} demonstrates that machine learning models outperform the benchmark econometric models in forecasting performance for the 1996-2009 period. During the 1997-1998 Asian financial crisis, the ridge regression (KRR) correlation model does not outperform other machine learning models, consistent with the results in Table \ref{tab:modelcompare5}, due to the small difference between extremes in the training data and prediction intervals. Overall, the predictive performance of the models in Table \ref{tab:ML1996-2009} aligns with the model performance for the 2005-2022 period presented in Tables \ref{tab:modelcompare4} and \ref{tab:modelcompare4}, indicating that the out-of-sample prediction results in this paper are robust and independent of the evaluation window time selection.

\begin{table}[]
\scriptsize
\begin{tabular}{c |c c c c| c c c c }
\hline
\multicolumn{1}{c}{\cellcolor[HTML]{FFFFFF}} & \multicolumn{4}{|c}{\cellcolor[HTML]{FFFFFF}RMSE} & \multicolumn{4}{|c}{\cellcolor[HTML]{FFFFFF}MAE} \\

\multicolumn{1}{c|}{\multirow{-2}{*}{Model}} & 1996-1999  & 1997-1998  & 2000-2003  & 2004-2009 & 1996-1999  & 1997-1998  & 2000-2003 & 2004-2009 \\
\hline
AR                                                                  & 0.88       & 0.88       & 1.02       & 1.58      & 0.72       & 0.76       & 0.86      & 1.41      \\
FM-AR-SE                                                            & 0.80       & 0.88       & 1.14       & 1.38      & 0.83       & 0.78       & 0.93      & 1.20      \\
XGB-GBTREE                                                          & 0.64       & 0.39       & 0.36       & 0.87      & 0.88       & 0.26       & 0.27      & 0.65      \\
XGB-GBLINEAR                                                        & 0.77       & 1.02       & 0.89       & 1.40      & 0.87       & 0.80       & 0.73      & 1.05      \\
GBDT-HUBER                                                          & 0.75       & 0.62       & 0.56       & 1.16      & 0.87       & 0.39       & 0.45      & 0.86      \\
GBDT-AE                                                             & 1.05       & 1.08       & 0.70       & 1.02      & 0.78       & 0.75       & 0.43      & 0.81      \\
GBDT-SE                                                             & 0.66       & 0.37       & 0.74       & 0.93      & 0.90       & 0.26       & 0.65      & 0.64      \\
RF-AE                                                               & 0.67       & 0.44       & 0.58       & 0.92      & 0.86       & 0.35       & 0.43      & 0.71      \\
RF-SE                                                               & 0.53       & 0.39       & 0.60       & 0.92      & 0.92       & 0.33       & 0.45      & 0.70      \\
FM-XGB-GBLINEAR                                                     & 0.81       & 0.85       & 1.22       & 1.22      & 0.85       & 0.74       & 0.98      & 1.00      \\
FM-XGB-GBTREE                                                       & 0.55       & 0.39       & 0.91       & 1.32      & 0.91       & 0.29       & 0.69      & 1.08      \\
FM-GBDT-AE                                                          & 0.67       & 0.38       & 0.81       & 1.18      & 0.91       & 0.28       & 0.60      & 1.00      \\
FM-GBDT-HUBER                                                       & 0.38       & 0.32       & 0.87       & 1.21      & 0.96       & 0.23       & 0.68      & 0.95      \\
FM-GBDT-SE                                                          & 0.44       & 0.44       & 0.94       & 1.30      & 0.96       & 0.37       & 0.73      & 1.04      \\
FM-RF-AE                                                            & 0.56       & 0.52       & 0.89       & 1.15      & 0.91       & 0.41       & 0.75      & 0.96      \\
FM-RF-SE                                                            & 0.54       & 0.61       & 0.79       & 1.13      & 0.91       & 0.46       & 0.65      & 0.95      \\
FM-KRR-POLY                                                         & 0.77       & 0.79       & 1.04       & 1.38      & 0.87       & 0.71       & 0.82      & 1.16      \\
FM-KRR-RBF                                                          & 0.77       & 0.80       & 1.05       & 1.39      & 0.87       & 0.72       & 0.83      & 1.17      \\
FM-LASSO                                                            & 1.15       & 1.11       & 0.98       & 1.62      & 0.87       & 1.01       & 0.79      & 1.39      \\
Mean G2                                                         & 0.54       & 0.44       & 0.51       & 0.88      & 0.42       & 0.36       & 0.38      & 0.70      \\
Median G2                                                       & 0.53       & 0.38       & 0.52       & 0.85      & 0.36       & 0.30       & 0.40      & 0.64      \\
Mean G3                                                         & 0.57       & 0.54       & 0.79       & 1.13      & 0.49       & 0.48       & 0.64      & 0.96      \\
Median G3                                                       & 0.54       & 0.53       & 0.84       & 1.21      & 0.44       & 0.44       & 0.68      & 1.00      \\
Mean G2 \& G3                                                         & 0.53       & 0.47       & 0.66       & 0.98      & 0.45       & 0.42       & 0.53      & 0.81      \\
Median G2 \& G3                                                        & 0.47       & 0.41       & 0.69       & 0.99      & 0.37       & 0.35       & 0.55      & 0.81      \\
RECIP\_4                                                            & 0.47       & 0.48       & 0.63       & 0.96      & 2.81       & 0.43       & 0.51      & 0.77      \\
RECIP\_6                                                            & 0.50       & 0.52       & 0.62       & 0.96      & 4.03       & 2.87       & 0.50      & 0.79      \\
RECIP\_8                                                            & 0.52       & 0.56       & 0.63       & 0.97      & 5.06       & 4.92       & 0.51      & 0.80      \\
EXP\_0.5\_4                                                         & 0.48       & 0.49       & 0.66       & 0.98      & 2.82       & 0.44       & 0.53      & 0.81      \\
EXP\_0.8\_4                                                         & 0.48       & 0.49       & 0.65       & 0.98      & 2.82       & 0.44       & 0.53      & 0.81      \\
EXP\_0.9\_4                                                         & 0.48       & 0.49       & 0.65       & 0.98      & 2.82       & 0.44       & 0.53      & 0.80      \\
EXP\_1\_4                                                           & 0.48       & 0.49       & 0.65       & 0.97      & 2.82       & 0.44       & 0.53      & 0.80      \\
EXP\_0.5\_6                                                         & 0.51       & 0.54       & 0.66       & 0.98      & 4.04       & 2.88       & 0.54      & 0.81      \\
EXP\_0.8\_6                                                         & 0.51       & 0.54       & 0.66       & 0.98      & 4.04       & 2.88       & 0.53      & 0.81      \\
EXP\_0.9\_6                                                         & 0.51       & 0.54       & 0.65       & 0.98      & 4.04       & 2.88       & 0.53      & 0.81      \\
EXP\_1\_6                                                           & 0.51       & 0.54       & 0.65       & 0.98      & 4.04       & 2.88       & 0.53      & 0.81      \\
EXP\_0.5\_8                                                         & 0.54       & 0.60       & 0.66       & 0.99      & 5.07       & 4.94       & 0.54      & 0.82      \\
EXP\_0.8\_8                                                         & 0.53       & 0.60       & 0.66       & 0.98      & 5.07       & 4.94       & 0.54      & 0.81      \\
EXP\_0.9\_8                                                         & 0.53       & 0.59       & 0.66       & 0.98      & 5.07       & 4.94       & 0.53      & 0.81      \\
EXP\_1\_8                                                           & 0.53       & 0.59       & 0.66       & 0.98      & 5.07       & 4.94       & 0.53      & 0.81     \\
\hline
\end{tabular}
    \caption{Model Performance 1996-2009}
    \label{tab:ML1996-2009} 
\end{table}

\section{Conclusions}
This paper employs machine learning models, combined models, mixed models and weighted models to forecast China's quarterly GDP growth, with an analysis of the model performance and the interpretability of the machine learning models.

The main findings of this paper are as follows: First, the prediction accuracy of machine learning models models generally surpasses that of traditional econometric models. Second, when the economic growth is stable, the prediction accuracy of some machine learning models or combined models are often superior to expert forecasts. Third, during periods of economic fluctuation, if the fluctuation range is within the historical range of the training data, machine learning models can achieve accurate predictions. Fourth, at certain ``historical inflection points", especially when the economic fluctuation range exceeds the historical range of the training data, machine learning models can predict the inflection points, but the accuracy may be lower than that of expert predictions. The interpretability analysis of machine learning models helps to better understand the impact of economic drivers on macroeconomic forecasts and identify the effects of different variables on economic growth across various historical periods.


\bibliographystyle{apalike}
\bibliography{reference}



\appendix

\setcounter{equation}{0}
\numberwithin{equation}{section}
\setcounter{figure}{0}
\numberwithin{figure}{section}

\renewcommand{\thesection}{Appendix A.}
\section{Supplementary Materials}
\renewcommand{\thesection}{A}

In this appendix, we discuss in detail the machine learning models used in this paper, which include: regularized linear regression (RLR) model, kernel ridge regression (KRR) model, random forest (RF) model, gradient boosted tree (GBDT), extreme gradient boosting (XGBoost), and so on.

\subsection{Regularized Linear Regression}
Linear models have very strong explanatory power, but when the dimension of the feature variables used in the model is high and the number of observations is relatively small, the model is prone to overfitting problems. To mitigate the overfitting problem, an important class of treatments in machine learning is regularization. The classical regularized linear regression models are mainly Ridge Regression, Least Absolute Shrinkage Selection Operator (LASSO) and ElasticNet. The three regularized linear regression models are as follows:

(a) Ridge Regression:

\begin{equation}
\min_{\beta} \| y - X\beta \|_2^2 + \lambda \| \beta \|_2^2.
\end{equation}

(b) Least Absolute Shrinkage and Selection Operator (LASSO):

\begin{equation}
\min_{\beta} \| y - X\beta \|_2^2 + \lambda \| \beta \|_1.
\end{equation}

(c) ElasticNet:

\begin{equation}
\min_{\beta} \| y - X\beta \|_2^2 + \lambda (\rho \| \beta \|_2^2 + (1 - \rho) \| \beta \|_1).
\end{equation}

On the one hand, regularized linear regression requires that the squared error of the fit be small. On the other hand, the parameters of the model are restricted or regularized by shrinking some of the parameters toward zero. The objective functions of all three types of regularized regression reflect a trade-off between the two requirements; the difference lies in the specific form of the trade-off between the two. Ridge regression requires a smaller number of norms for the model parameters, while LASSO requires a smaller number of norms for the model parameters, and the elasticity network combines the characteristics of both ridge regression and LASSO.

\subsection{Kernel Ridge Regression}
Kernel Ridge Regression (KRR) combines the two methods of ridge regression and kernel technique, which belongs to a kind of nonlinear regression. Since the prediction function of ridge regression is linear, if the potential mapping between the explanatory variables and the explained variables is linear, the prediction effect of ridge regression is very good. However, in reality, many mapping relationships are often nonlinear, and the kernel method provides a nonlinear learning model. The basic idea of the kernel method is to embed the original data into a suitable high-dimensional feature space through some nonlinear mapping, and utilize a linear learner to learn in the new feature space. Kernel methods transform a nonlinear problem in the input space into a linear learning problem in a high dimensional feature space.

Let x be the feature mapping from the input space X to the feature space H. Define the kernel function as $K(x_1, x_2) = \langle \phi(x_1), \phi(x_2) \rangle_{\mathcal{H}}$. Kernel ridge regression can be expressed as the following problem:

\begin{equation}
\min_{\beta} \| y - \phi(X) \beta \|_2^2 + \lambda \| \beta \|_2^2.
\end{equation}

In solving real-world problems, feature mappings are often unknown in advance, so the construction and computation of feature mappings require a high level of experience and skill. Kernel techniques use kernel functions to solve optimization problems directly, which can circumvent the tedious solution process of feature mappings and thus accelerate the computation of kernel methods. Commonly used kernel functions include polynomial kernel and Gaussian kernel.

\subsection{Random Forest}

Random Forest (RF) is a class of non-parametric learning methods that combine the results of multiple base-learner decision trees through the idea of integrated learning to obtain more accurate and stable predictions. The method compensates for the inherent flaws of a single model by integrating a large number of base learners that have learned and predicted independently of each other, and complementing their strengths and weaknesses to make better predictions. The single base-learner decision tree used in random forests is a tree structure that actually divides the input space with hyperplanes, each division dividing the current space into two, and eventually dividing the input space into multiple small subregions. The mathematical expression of the decision tree function is as follows:

\begin{equation}
f(x) = \sum_{m=1}^{M} c_m I(x \in R_m),
\end{equation}
where the input space is partitioned into $M$ copies, $R$ is one of the subregions $I(\cdot \in R_m)$, and I is the schematic function of the subregion. Under squared error loss, $c_m = \text{avg}(y_i \mid x_i \in R_m)$.

Random Forest generates a large number of decision trees in parallel in a randomized manner and seeks to make the correlation between each decision tree small enough to enable each decision tree to make predictions relatively independently. In order to reduce the dependence between decision trees, Random Forest mainly starts from the random selection of samples and the random selection of features. Bootstrap randomly select a sample from the sample set, and then randomly select a sub-feature from all the features, and select the best split feature as a node to generate a decision tree. Repeat the above method to generate a large number of relatively independent decision trees. By averaging the prediction results of these decision trees, the comprehensive prediction of the whole random forest is obtained. 

Random forests have many advantages of being able to combine features and thus learn non-linear structures such as interactions between features. By introducing randomness in sample and feature selection, etc., and integrating a large number of relatively independent decision trees, the overall model has good resistance to overfitting and very good resistance to noise, and the prediction results are very stable. The mutual independence of each decision tree also makes the model easy to parallelized the computation and has a fast training speed.

\subsection{Gradient Boosting Decision Tree}
Gradient Boosting Decision Tree (GBDT) is also an integrated learning method that uses a decision tree as a base learner. Unlike the parallel approach of Random Forest where each decision tree is relatively independent, Gradient Boosting Decision Tree (GBDT) is a class of machine learning methods that uses an additive model and a forward stepwise algorithm for gradient boosting.GBDT is an iterative decision tree algorithm that linearly sums up a large number of decision trees into an additive model in series, so that the conclusions from all the decision trees are summed up to obtain the final prediction result. The mathematical expression of the gradient boosting tree model is as follows:

\begin{equation}
f_M(x) = \sum_{m=1}^{M} T_m(x; \Theta_m),
\end{equation}
where $ T_m(x; \Theta_m)$ is the decision tree generated at step m, $\Theta_m$ is the parameters of that decision tree, and M is the number of decision trees.

The learning of the prediction function requires solving the following optimization problem:
\begin{equation}
\min_{\Theta} L(y, f_M(x)) = \min_{\Theta} L \left( y, \sum_{m=1}^{M} T_m(x; \Theta_m) \right).
\end{equation}

The problem is transformed into the following iterative process:
\begin{equation}
f_m(x) = f_{m-1}(x) + T_m(x; \Theta_m).
\end{equation}

The gradient boosting algorithm requires M iterations, each iteration generating a decision tree such that the loss function of the decision tree generated in each iteration is minimized for the training set. When the loss function is squared error, each iteration actually learns based on the residuals of the model obtained in the previous step. When the loss function is not a squared error, Freidman proposed the gradient boosting algorithm, which utilizes the negative gradient of the loss function under the current model as an approximation of the residuals to generate the decision tree for the next step. The loss function is made smaller and smaller by this iteration. For different gradient boosting trees, the main difference is the loss function. For regression problems, the loss functions are generally squared error, absolute error and Huber error loss.

The main disadvantage of GBDT is mainly that it is difficult to train in parallel because the base learners are in a serial process with each other.

\subsection{Extreme Gradient Boosting}

Extreme Gradient Boosting (XGBoost) is developed from GBDT, which is also an integrated learning method based on decision trees and utilizes an additive model and a forward-stepping algorithm for learning optimization.

The difference between XGBoost and GBDT is mainly manifested in the two aspects of the objective function and optimization method.The regularization term to control the model complexity is added to the objective function of XGBoost in order to prevent the overfitting phenomenon from occurring. Specifically, its objective function is:

\begin{equation}
\min_{\Theta} L(y, f_M(x)) + \sum_{m=1}^{M} \Omega(T_m),
\end{equation}
where $\Omega(T_m) = \gamma |T_m| + \frac{1}{2} \lambda \|W_m\|_2^2$ is the model complexity of each decision tree, $\gamma$ and $\lambda$ are hyperparameters, $\left|T_m\right|$ is the number of leaf nodes, and $w_m$ is the leaf node weight vector. In the optimization method, GBDT uses only the first-order gradient information in the optimization, while XGBoost performs a two-order Taylor expansion of the loss function in the optimization, and uses both the first-order and second-order derivative information, which makes the accuracy higher.

In addition, the optimization process of XGBoost can be parallelized to support distributed training on multiple machines, whereas GBDT has no parallelization design. Parallelization is not the simultaneous computation of decision trees, each decision tree training still needs to wait for the previous step of decision tree training to complete before starting training. XGBoost's parallelization is mainly in the feature dimension, using a parallel way to find the best segmentation point of each feature, which greatly improves the speed of training. In terms of model flexibility, GBDT uses decision trees as base classifiers, while XGBoost not only supports decision trees, but can also support multiple types of base learners such as linear learners.

\end{document}